\documentclass[twocolumn]{aastex631}
\usepackage{multirow}
\usepackage{graphics}

\newcommand{\degree}{$^\circ$}

\shorttitle{Atmospheric Characterization of Hot Jupiter CoRoT-1 b}
\shortauthors{Glidic et al.}

\begin{document}

\title{Atmospheric Characterization of Hot Jupiter CoRoT-1 b Using the Wide Field Camera 3 on the Hubble Space Telescope}

\correspondingauthor{Kayli Glidic}
\email{kayliglidic@email.arizona.edu}
\email{kayliglidic@gmail.com}

\author[0000-0003-4669-7088]{Kayli Glidic}
\affiliation{Steward Observatory, The University of Arizona, Tucson, AZ 85721, USA}

\author[0000-0001-8291-6490]{Everett Schlawin}
\affiliation{Steward Observatory, The University of Arizona, Tucson, AZ 85721, USA}

\author[0000-0002-3295-1279]{Lindsey Wiser}
\affiliation{School of Earth \& Space Exploration, Arizona State University, Tempe, AZ 85287, USA}

\author[0000-0003-2969-6040]{Yifan Zhou}
\altaffiliation{51 Pegasi b Fellow}
\affiliation{Department of Astronomy, The University of Texas at Austin, Austin, TX 78712, USA}

\author[0000-0001-5727-4094]{Drake Deming}
\affiliation{Department of Astronomy, University of Maryland at College Park, College Park, MD 20742, USA}

\author[0000-0002-2338-476X]{Michael Line}
\affiliation{School of Earth \& Space Exploration, Arizona State University, Tempe, AZ 85287, USA}
 
\begin{abstract}
Exoplanet CoRoT-1 b is intriguing because we predict it to be a transitional planet between hot Jupiters (equilibrium temperatures $\sim$ 1500 K) and ultra-hot Jupiters (equilibrium temperatures $>$ 2000 K). In 2012, observations of CoRoT-1 b included one primary transit and three secondary eclipses with the Hubble Space Telescope (HST) Wide Field Camera 3 (WFC3) combined with the G141 grism ($1.1-1.7$ $\micron$) in stare mode. 
We aimed to further investigate CoRoT-1 b through its secondary eclipses, producing spectrophotometric light curves corrected for charge trapping, also known as the ramp effect in time-series observations with the WFC3. We found that, when correcting for the ramp effect and using the typically discarded first orbit, we are better capable of constraining and optimizing the emission and transmission spectra. We did a grid retrieval in this transitional temperature regime and found the 
spectra for CoRoT-1 b to be featureless and to agree with an inverted temperature–pressure (T–P) profile. We note, however, that the contribution function for the WFC3 indicates pressures probed near $10^{-3}$ to $10^{0}$ bar, which correspond to a nearly isothermal region in our T–P profile, thereby indicating that the inversion at high altitude is model-dependent. Despite no distinct features, the analysis done on CoRoT-1 b paves the way to high-precision results with stare mode spectroscopy. As a new generation of observations from the James Webb Space Telescope (JWST) approaches, CoRoT-1 b might be an interesting follow-up target because the time-series spectroscopic modes of JWST's NIRSpec, MIRI, and NIRCam instruments will be analogous to HST's stare mode.
\end{abstract}

\keywords{\href{https://vocabs.ands.org.au/repository/api/lda/aas/the-unified-astronomy-thesaurus/current/resource.html?uri=http://astrothesaurus.org/uat/487}{Exoplanet atmospheres (487)}; \href{http://astrothesaurus.org/uat/498}{Exoplanets (498)}; \href{https://vocabs.ands.org.au/repository/api/lda/aas/the-unified-astronomy-thesaurus/current/resource.html?uri=http://astrothesaurus.org/uat/442}{Eclipses (442)}; \href{https://vocabs.ands.org.au/repository/api/lda/aas/the-unified-astronomy-thesaurus/current/resource.html?uri=http://astrothesaurus.org/uat/1558}{Spectroscopy (1558)}; \href{https://vocabs.ands.org.au/repository/api/lda/aas/the-unified-astronomy-thesaurus/current/resource.html?uri=http://astrothesaurus.org/uat/761}{Hubble Space Telescope (761)}; \href{https://vocabs.ands.org.au/repository/api/lda/aas/the-unified-astronomy-thesaurus/current/resource.html?uri=http://astrothesaurus.org/uat/753}{Hot Jupiters (753)}}

\section{Introduction} \label{sec:Introduction}
Despite the inability to spatially resolve transiting exoplanets from their host stars, their atmospheres contain a wealth of information that has subsequently placed priority on the spectroscopic analysis of atmospheres and the development of interpretive tools. The spectroscopy of exoplanets can be achieved with several techniques but requires a geometrically favorable system. These transiting exoplanets present a unique insight into their atmospheres. During a primary transit, a small fraction of stellar flux is blocked and filtered through the planet's atmosphere, while, during a secondary eclipse, the planet's thermal radiation and reflected light disappear and then reappear. For a detailed review see \citet{Winn_exoplanet} and \citet{kreidberg2018review}. Probing these environments has returned numerous insightful measurements on wavelength-dependent transit and eclipse depths, unveiling atmospheric chemical compositions and thermal structures. The interpretations of these results can further yield the physics and chemistry that govern these atmospheres alongside feasible constraints on the formation, evolution, and climate of these exoplanets \citep[e.g.][]{oberg11,kreidberg2014wasp43,mordasini2016planetFormationSpec,sing2016continuum,espinoza2017metalEnrichmentCtoO,baxter2020}.
 
In the spectroscopy of exoplanets, many species in the atmospheres of these planets are potentially detectable due to strong absorption features. In particular, with hot Jupiters, sodium and potassium are dominant absorbers in the optical wavelength range; water and carbon monoxide are dominant absorbers in the near-infrared wavelength range \citep{kreidberg2018review}. Transmission and emission spectra have already detected Na \citep{Charbonneau2002-Na, Nikolov2014-Na}, K \citep{Sing2011-K}, H \citep{vidalmadjar, ehrenreich2015hydrogen}, and $H_2O$ \citep{deming13, huitson13, birkby13, fraine2014hatp11, kreidberg2014wasp43, kreidberg2015wasp12, McCullough2014-h20, line2016hd209wfc3, evans2016wasp121H2OTiO, stevenson2016hatp26}. The water absorption feature at 1.4~$\micron$ is the most commonly observed molecular species, with most spectra coming from the Hubble Space Telescope (HST) Wide Field Camera (WFC3) G141 grism operating in the near-infrared (1.1 - 1.7 $\micron$). Furthermore, the transmission and emission spectra can also constrain absolute abundance and abundance ratios for these species in the atmospheres of exoplanets. One particular atmospheric abundance ratio to consider is the carbon-to-oxygen ratio, C/O, a critical ratio to trace formation pathways \citep{oberg11,madhusudhan12, madhsudhan2016reviewAtm}.

These techniques favor large, hot gas giants classified as hot Jupiters. Many of the mentioned molecular detections have been from hot Jupiters, a planetary class characterized by short orbital periods, proximity to their host stars, high temperatures, and a similar mass to Jupiter, and they are relatively easy to detect \citep{fortney2021review}. Hot Jupiters are usually tidally locked, having a synchronous rotation of the planet's orbit and spin, with a permanent dayside and permanent nightside. These tidally locked hot Jupiters allow for the measurement of dayside thermal flux determined by the temperature–pressure (T–P) profile and atmospheric opacities \citep{baxter2020}. Typical hot Jupiters have an equilibrium temperature around 1500 K with a non-inverted T–P profile: the temperature decreases with altitude. Ultra-hot Jupiters are thought to exhibit different atmospheric properties with equilibrium temperatures $>$ 2000 K and may have inverted T–P profiles: the temperature increases with altitude \citep{baxter2020}. For example, in HST WFC3 observations of WASP-33 b, the best-fit modeled T–P profile was inverted around a pressure of $\sim$0.1 bar \citep{2015WASP33b}. Temperature inversions are thought to be a result of the presence of titanium oxide and vanadium(II) oxide (TiO/VO), which are expected to be found in the hottest of atmospheres, or other absorbers of stellar radiation \citep{deming11}. Even in small amounts, these elements are strong absorbers capable of heating the upper atmospheres of these hot Jupiters \citep{kreidberg2018review}.

Hot Jupiter CoRoT-1 b, a discovery reported in \citet{barge08}, orbits a metal-poor G0V star of magnitude \textit{V} = 13.6 and effective temperature \textit{T} = 5950 K in a period of about 1.5 days. CoRoT-1 b is notably interesting as we predict it to be in the transitional realm between hot Jupiters and ultra-hot Jupiters, with an initially reported equilibrium temperature of $1898\pm50$ K \citep{barge08}. In \citet{deming11}, the Warm Spitzer mission with the Infrared Array Camera (IRAC) observed secondary eclipses of CoRoT-1 b at 3.6 and 4.5 $\micron$, with observations of stronger eclipses at 4.5 $\micron$, often indicative of an atmospheric temperature inversion. The multiband photometry found matched a 2460 K blackbody, a possible indication of a high-altitude layer that strongly absorbs stellar irradiance or an isothermal temperature gradient \citep{deming11}. 

We organize the paper as follows. We describe our HST observations in section \ref{sec:Observations}. Section \ref{sec:Data Reduction} details methods for our data reduction, correction of the systematic ramp effect, and optimization of light curve fits. In section \ref{Results_Discussion}, we present our modeled light curve fits with their corresponding statistically optimal parameters and resulting spectra. Additionally, we describe our analysis of CoRoT-1 b's properties based on atmospheric grid retrievals for the emission spectrum and comparisons with previously published results for the transmission spectrum. Finally, we conclude and note the potential benefits of a follow-up with the James Webb Space Telescope (JWST) in section \ref{sec:Conclusion}. 

\section{Observations} \label{sec:Observations}
We utilized time-series spectroscopy during one primary transit and three secondary eclipses of CoRoT-1 b between UT 2012 January 17 and February 5 with the Wide Field Camera 3 (WFC3) infrared (IR) detector aboard the Hubble Space Telescope (HST) as part of GO Program 12181. While the single primary transit was published previously by \citet{ranjan2014ApJ...785..148R}, the three secondary eclipses were not. 

The IR channel of the WFC3 is a 1K $\times$ 1K HgCdTe detector with an assemblage of 17 filters and two grisms. All four visits of CoRoT-1 b used grism G141, which is widely used for exoplanet spectroscopy \citep[e.g.][]{wakeford2016marginalizingSys} and spans 1.1–1.7 $\micron$\ at \textit{R} = $\lambda/\Delta\lambda$ $\approx$  130. Each visit consisted of four HST orbits of $\sim$96 minutes with $\sim$51 minutes of observing time and $\sim$45 minute data gaps between orbits due to Earth occultation. At the beginning of each visit, a direct image was taken with the F139M filter for wavelength calibrations following the method outlined in \citet{2014ApJ...783..113W}. The spectroscopic data collected were obtained in stare mode prior to the implementation of the popular technique of spatially scanning the telescope \citep{mccullough2012spatialScan} with the 128 $\times$ 128 subarray, using the SPARS10, NSAMP = 16 readout pattern, which corresponds to an exposure time of 100.651947 s. Table \ref{tab:observations_summary} summarizes our observations. 
\vspace{1.0cm}
\begin{deluxetable}{ccccc}[htb!]
\tablenum{1}
\tablecaption{Summary of Observations for CoRoT-1 b \label{tab:observations_summary}}
\tablewidth{0pt}
\tablecolumns{4}
\tablehead{
\multicolumn{1}{c}{Visit}& \multicolumn{1}{c}{Number} & \multicolumn{1}{c}{Start Date (UT)} & \multicolumn{1}{c}{Orbital Phase} &\\
\multicolumn{1}{c}{Number} & \multicolumn{1}{c}{of Orbits} & \multicolumn{1}{c}{2012} & 
\multicolumn{1}{c}{Range}
}
\startdata
1* & 4 & January 23 & 0.028 - 0.876\\
2 & 4 & January 17 & 0.379 - 0.531\\
3 & 4 & January 27 & 0.388 - 0.542\\
4 & 4 & February 5 & 0.384 - 0.537
\enddata
\tablecomments{$^*$ We denote the primary transit as visit number 1 despite the observation date.}
\end{deluxetable}

\section{Data Reduction} \label{sec:Data Reduction}
\subsection{Spectral Extraction} \label{sec:Spectral Extraction}

We reduced the HST WFC3 data utilizing a spectroscopic data-reduction pipeline with the \texttt{Time Series Helper and Integration Reduction Tool (tshirt)}, \footnote{\texttt{tshirt} website: \url{https://tshirt.readthedocs.io/en/latest/}} which is a general-purpose tool for time-series science. We begin with \texttt{flt} image files generated from the \texttt{calwf3} pipeline. These \texttt{flt} image files are biased-subtracted, dark-current-corrected, cosmic-ray-rejected, and gain-calibrated. The science images are also corrected for the nonlinear response of the detectors, flagging pixels that extend into saturation to compute the count rate per pixel. Therefore, each pixel is in units of electrons/second. We also converted the Modified Julian Date (MJD) = JD $-$ 2,400,000.5 timestamp on the images to Barycentric Julian Date (BJD) to correct for differences in the Earth’s orbital position with respect to the barycenter of the solar system.

\subsection{Background Subtraction} \label{Background Subtraction}

\begin{figure}[htb!]
    \centering
    \includegraphics[width=0.45\textwidth]{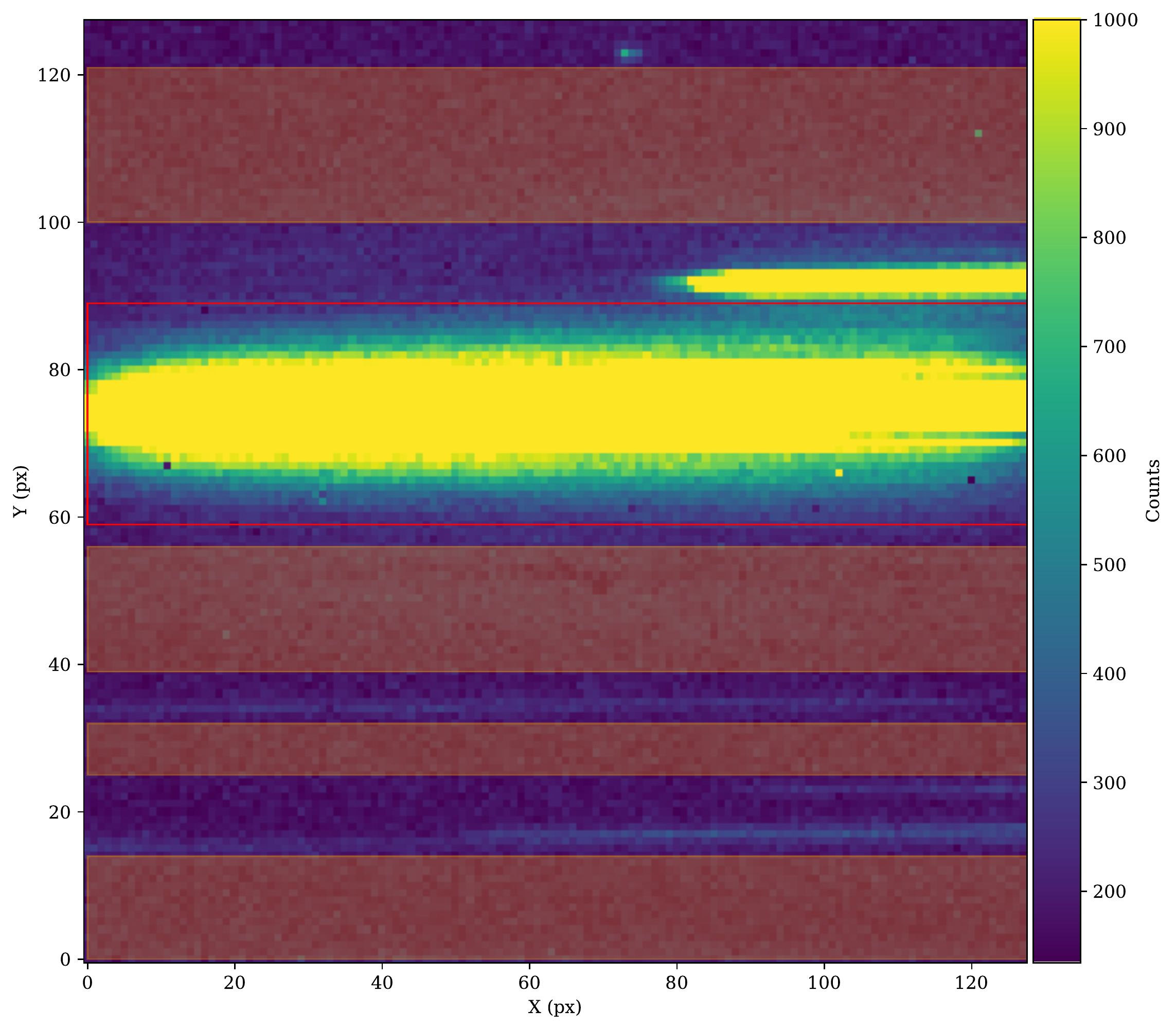}
    \caption{Image of CoRoT-1 b, Visit 2, displayed in relation to its location on the instrument detector (in pixels). The color bar depicts the count value at each pixel. The red rectangle outlining the source, CoRoT-1 b, represents the source aperture, the spectral extraction area. The shaded red rectangular regions represent the areas for background subtraction. The varying widths for the background subtraction regions are to avoid neighboring sources of flux.}
    \label{fig:starlabel}
\end{figure}

As our rectangular source aperture includes some background flux, and as the background flux varies with time, subtraction of the background from our time-series spectra is required to produce an accurate light curve. Our background estimation areas were rectangular with the same center position of dispersion (\textit{x}-value) as the source aperture and matched the wavelength range of the spectrum. We identified these background estimation areas by choosing areas on the detector free of object flux and neighboring objects in each 2D \texttt{flt} image file as seen in Figure \ref{fig:starlabel}. We subtracted in the spatial direction (\textit{y}-direction) column-by-column with a linear fit. The rectangular background subtraction areas varied in width and number between visits, based on flux sources present in the background, to reduce residuals in the light curves. Figure \ref{fig:starlabel} shows an image of CoRoT-1 b with the area of spectral extraction outlined in red and the areas of background subtraction shaded in red. Therefore, repeating this process of background subtraction for every \texttt{flt} image file (with the same parameters for every file for a particular visit number) produces a time-series with background-subtracted source flux.

\subsection{Spectral Image Alignment} \label{Spectral Image Alignment}

We aligned all \texttt{flt} image files along the \textit{x}-axis corresponding to wavelength. This alignment was achieved by first filtering the spectra with a digital high- and low-pass filter and then cross-correlating the results to find the sub-pixel shift value. For the high-pass filter, we use a \texttt{scipy} Butterworth signal filter with a bandpass from 0.05 to 0.55 of the Nyquist frequency. The maximum frequency removes any pixel-to-pixel flat-field residuals, and the minimum frequency flattens the spectrum to ensure the cross-correlation peak is well defined. After locating the cross-correlation peak, the spectra were linearly interpolated onto a common grid set by a reference spectrum in the middle of the time-series. Without the alignment of the \texttt{flt} image files, spikes in the flux might have been noticeable at the start of visits because of small shifts in the \textit{x}-direction. These are most noticeable where the slope of the spectrum is steepest.

\subsection{Wavelength-dependent Time-series} \label{Wavelength-dependent Time Series}

We extracted all spectroscopic data from the \texttt{flt} image files, accounting for background subtraction, utilizing \texttt{tshirt}'s optimal extraction method \citep{horne1986optimalE} in order to create a wavelength-dependent time-series. We binned our time-series into ten wavelength channels. We binned the time-series by wavelength to analyze the light curves in smaller regions of the wavelength range to increase the signal-to-noise ratio, decrease the number of free parameters in our model, and produce a transmission spectrum and emission spectra for CoRoT-1 b. Table \ref{tab:secondary_eclipse_parameters} and Table \ref{tab:primary_eclipse_parameters} contain the wavelength boundaries corresponding to all visits for each bin.

\subsection{Correction of Systematics} \label{Correction of Systematics}

In time-resolved observations with the WFC3 IR channel, the ``ramp effect" is a dominant source of systematics, limiting photometric precision \citep{Deming2006,berta2012flat_gj1214,zhou2017chargeTrap}. The ramp effect is an exponentially shaped signal in the time domain more severe in the first orbit with HST data. Charge trapping, attributed as the major cause of the ramp effect, is due to defects in near-IR detectors \citep{SmithHgCdTeTheory,SmithHgCdTeCalibration,zhou2017chargeTrap}. At the beginning of observations, a fraction of the stimulated charges is trapped in these defects, leading to reduced flux measurements. As the traps fill, the ramp profile flattens. Now, the systematics between orbits is not fully understood. Pointing the telescope at a new target, changing the insolation of the spacecraft, and beginning to fill charge traps, it takes time for the instrumentation to settle into a repeatable configuration. Consequently, this leads to discarding the first orbit in many visits, and many use a divide-out-of-transit method with an exponential model \citep{berta2012flat_gj1214}. Instead, this paper will focus on the use of the \texttt{Ramp Effect Charge Trapping Eliminator (RECTE)}, described in \citet{zhou2017chargeTrap} as a physically motivated model that takes into account the number of charge carrier traps, the trapping efficiency, and the trap lifetimes of every pixel on the detector. \citet{zhou2017chargeTrap} developed \texttt{RECTE} to model the charge trapping procedure. This model provides a theoretical ramp profile, which is a function of the detector irradiation level, the initial state of the WFC3 detector, and the additional charges trapped during the Earth occultation. We used \texttt{RECTE} to study all orbital data on CoRoT-1 b and to analyze its transmission spectrum and emission spectra.  

Initially, we plotted the extracted light curves following background subtraction, but before dividing out any systematic ramps. As seen in the leftmost panel of Figures \ref{fig:lightcurve_visit2}, \ref{fig:lightcurve_visit3}, \ref{fig:lightcurve_visit4}, and \ref{fig:lightcurve_visit1}, the characteristic ramp is visible in each visit’s raw light curves, with the ramps on the first orbits of each observation having greater amplitudes. These ramp profiles also vary between wavelength channels as a result of the different charge trapping rates from variations in flux intensity. As \citet{zhou2017chargeTrap} explains in further detail, \texttt{RECTE} allows one to quantitatively model the charge carrier trapping process on a channel-to-channel basis. This ability allows one to precisely calibrate the ramp-effect-impacted time-resolved observations made with the WFC3 as opposed to fitting empirically derived exponential functions.

\texttt{RECTE} can be applied to both scanning and stare mode observations but with notable differences in the application. In stare mode, used in our observations of CoRoT-1 b, pixels are illuminated at vastly different levels, producing vastly different count levels not reflected by an average. Illumination levels are a critical factor in determining the ramp profile. At high illumination levels, traps fill faster, and the ramp rises faster, and vice versa. The large variations between pixels mean that each pixel will have a significantly different ramp profile. Therefore, one must loop through each pixel and model it individually to calculate the combined ramp profile in stare mode. We determined the illumination level for CoRoT-1 b by calculating the median \texttt{fits} file. From the 128 $\times$ 128 pixel subarray, we implemented a bounding-box around the source, for \texttt{RECTE} calculations, from 0 to 128 pixels in the \textit{x}-direction and 59 to 89 pixels in the \textit{y}-direction, a total of 3840 pixels involved. We calculated the charge trap evolution for each pixel independently and did an aperture sum of the bounding-box pixels in the spatial direction. When applying \texttt{RECTE}, there are four degrees of freedom (see Section \ref{sec:Optimizing Light Curve Fits}) that deal with the initial number of charge carrier traps and the additional number of charge carriers trapped during the gaps between orbits \citep{zhou2017chargeTrap}. For a detailed description and correction, see \citet{zhou2017chargeTrap}. For each trap population, the charge trapping process is controlled by the total number of traps (\texttt{nTrap\_s} = 1525.38 traps, \texttt{nTrap\_f} = 162.38 traps), the trapping efficiencies [unitless] (\texttt{eta\_s} = 0.013318, \texttt{eta\_f} = 0.008407), and the trap lifetimes/timescales (\texttt{tau\_s} = 16.3 ks, \texttt{tau\_f} = 281.463 s). The parameter names in parenthesis are defined and have fixed values as used in \citet{zhou2017chargeTrap}. Additionally, we assume that all pixels within each wavelength channel share the same initial numbers of trapped charges and additional charges trapped during Earth occultations. The illumination level of the pixels determines their individual ramp profiles. Since we are unaware of the precise charge trap values (initial and additional charges), we optimized these values simultaneously with transit-related parameters when applying light curve fits for both the transit and eclipses of CoRoT-1 b.

\subsection{Spitzer Reanalysis} 
\label{Spitzer Re-analysis}
Spitzer secondary eclipse depths for CoRoT-1 b were published by \citet{deming11}.  Those authors used a simple quadratic function of \textit{X}- and \textit{Y}-positions of the star on the detector to decorrelate Spitzer's intrapixel photometric variations. In the past decade, significant advances have been made in techniques used to remove Spitzer's intrapixel effect and (sometimes) temporal ramps (e.g., \citealp{stevenson2012, deming2015secE, ingalls2016, garhart2020}). We have therefore updated CoRoT-1 b's Spitzer eclipse depths. One of us (D.D.) is pursuing a large population study of hot Jupiter emission that involves a comprehensive reanalysis of all Spitzer secondary eclipse data for gaseous exoplanets. We adopted updated eclipse depths from that study. It uses a pixel-level-decorrelation methodology \citep{deming2015secE}, and codes identical to those used by \citet{garhart2020}. One difference is that the population study adopts slightly different orbital parameters for CoRoT-1 b than in our Table \ref{tab:fixedparams} (e.g., it uses $i=83.88^{o}$ and $a/R = 4.751$).  We performed some additional analyses of Spitzer's CoRoT-1 data to verify that the small difference in orbital parameters does not significantly affect the eclipse depths. Our new eclipse depths are $4602\pm477$ and $4380\pm440$\,ppm at 3.6 and 4.5\,$\mu$m, respectively. Both values agree with the original values from \citet{deming11} to within $1\sigma$.

\subsection{Optimizing Light Curve Fits} \label{sec:Optimizing Light Curve Fits}

The basis of the light curve fitting process jointly implements \texttt{RECTE} and \texttt{\textbf{BA}sic \textbf{T}ransit \textbf{M}odel c\textbf{A}lculatio\textbf{N} (batman)}, a Python package for modeling exoplanet transit and secondary eclipse light curves outlined in \citet{batman}.While \texttt{batman} is capable of modeling transit and secondary eclipse light curves based on given planet parameters, it does not account for the ramp systematic or the linear trend in the baseline present in the data. Therefore, we scaled the modeled flux from \texttt{batman} by the ramp profile calculated with \texttt{RECTE} and a linear regression as follows:

\vspace{0.5cm}
\begin{center}
\begin{math}
F_{modeled} = F_{\texttt{batman}}(t, \delta)\times F_{\texttt{RECTE}}(t, \vec{T}) \times (a+bX),
\end{math}
\end{center}
\vspace{0.5cm}
where $F_{modeled}$ is the scaled modeled flux used in our light curve fits. $F_{\texttt{batman}}$ is the initial modeled flux from \texttt{batman}, which depends on time \textit{t} and $\delta$, a variable for \texttt{fp} (or interchangeably \texttt{rp}). $F_{\texttt{RECTE}}$ is the flux of the ramp profile, which depends on time \textit{t} and $\vec{T}$, a vector containing the charge trapping parameters: \texttt{trap\_pop\_s, trap\_pop\_f, dtrap\_s}, and \texttt{dtrap\_f}. Lastly, $(a+bX)$ is the linear regression applied to model a linear-in-time baseline slope where $X =  t - t_{minimum}$. See Table \ref{tab:freeparams_summary} for parameter definitions. 

\vspace{1cm}
\begin{deluxetable}{ccccc}[htb!]
\tablenum{2}
\tablecaption{Summary of Modeled Free Parameters \label{tab:freeparams_summary}}
\tablewidth{0pt}
\tablecolumns{2}
\tablehead{
\multicolumn{1}{c}{Free Parameter} & \multicolumn{1}{c}{Definition}
}
\startdata
 \texttt{fp} & The planet-to-star flux ratio\\
\texttt{rp} & The planet-to-star radius ratio \\
\hline
\texttt{a} & Linear regression \textit{y}-intercept \\
\texttt{b} & Linear regression slope\\
\hline
\texttt{trap\_pop\_s} & The initial number of occupied charge\\
& carrier traps for the slow population \\
\texttt{trap\_pop\_f} &  The initial number of occupied charge\\
& carrier traps for the fast population \\
\texttt{dtrap\_s} & Additional trapped charge carriers \\ & added in the middle of two orbits\\ & for slow population\\
\texttt{dtrap\_f} & Additional trapped charge carriers \\ & added in the middle of two orbits\\ & for fast population
\enddata
\tablecomments{All names of free parameters are defined as used in the Python script. For a more detailed description of how these parameters are used in the modeling process, see section \ref{sec:Optimizing Light Curve Fits}.}
\end{deluxetable}

We calculated fits for three secondary eclipses and one primary transit, each separated into ten wavelength channels. Typically, when fitting a transmission light curve model to the data, the transit depth, eclipse depth, and parameters relevant to the systematics modeled are allowed to vary. All other parameters tend to remain fixed because we do not expect them to vary across spectroscopic channels. Specific to CoRoT-1 b, the following parameters were fixed: orbital inclination (\textit{i}), period (\textit{P}), semi-major axis ($a/R_\star$) in units of stellar radius, eccentricity (\textit{e}, assuming circular orbit), longitude of periastron ($\omega$), and time of inferior conjunction ($T_c$). These values can be found in Table \ref{tab:fixedparams}. Modeling the secondary eclipses allowed the planet-to-star flux ratio (\texttt{fp}) to vary, while the primary transit allowed for the planet-to-star radius ratio (\texttt{rp}) and the limb darkening profile to vary. All fixed values of the models were obtained or calculated from values found in \citet{bonomo2017harpsMasses}. The systematics modeled in the light curve fits incorporated \texttt{RECTE} corrections to account for the ramp effect in the flux data. \texttt{RECTE} assumes two populations of charge traps distinguished by their trapping timescales. The slow trap population releases trapped charges on longer timescales and has a stronger effect overall, while the fast trap population has a stronger effect at the beginning of orbits. In total, there were seven free parameters considered within our models. The parameter names used in the Python script are in parentheses and short descriptions can be found in Table \ref{tab:freeparams_summary}. First, the planet-to-star flux ratio (\texttt{fp}), or interchangeably the planet radius (\texttt{rp}), depending on the transit type, is directly proportional to the eclipse depth (or transit depth as $(\frac{R_p}{R_\star})^2$) of the model and a value that varies across spectroscopic channels. Next, there is the \textit{y}-intercept (\texttt{a}) and the slope (\texttt{b}) of the linear regression we applied to the models to capture any overall linear trend in the normalized light curves, particularly in the baseline. Finally, there are the four \texttt{RECTE} charge trap parameters: the initial number of occupied charge carrier traps for the fast and slow populations (\texttt{trap\_pop\_s, trap\_pop\_f}) and the number of additional trapped charge carriers added in the middle of two orbits (\texttt{dtrap\_s, dtrap\_f}) for each population respectively. 

\begin{deluxetable}{ccccc}[htb!]
\tablenum{3}
\tablecaption{Star and Planet Parameters for CoRoT-1 b \label{tab:fixedparams}}
\tablewidth{0pt}
\tablecolumns{2}
\tablehead{
\multicolumn{1}{c}{Star Parameters} & \multicolumn{1}{c}{Value}
}
\startdata
$T_{eff}$ [K] & $5950 \pm 150$\\
log g & $4.311 \pm 0.019$\\
$R_\star$ $[R_\odot]$ & $1.131 \pm 0.045$\\
\hline
Planet Parameters & Value\\
\hline
\textit{i} [deg] & 85.15\degree \\
P [days] & 1.5089682\\
$a/R_\star$ &  4.811  \\
e & 0 \\
$\omega$  [deg] & 90\degree\\
$T_c$ [days] & 2454138.32807\\
\hline
Calculated Equilibrium\\ Temperatures [K] &
Day–Night Redistribution (\textit{f})\\
\hline
$1918$ & 1.0 (full)\\
$2123$ & 1.5\\
$2281$ & 2.0 (dayside-only)\\
$2450$ & 2.66 (weak/no redistribution)
\enddata
\tablecomments{Physical properties of the star and planet we obtained from \citet{corot1params} and \citet{bonomo2017harpsMasses}. Equilibrium temperature calculations for various day–night heat redistributions reference an equation similarly introduced in \citet{line2013chimera}. In \citet{line2013chimera}, the heat redistribution parameter is denoted as \textit{$\beta$}. The heat redistribution parameter \textit{f} that we use is specifically introduced in \citet{arcangeli2018wasp18}. All equilibrium temperature calculation assume zero albedo.}
\end{deluxetable}

As the exact values for the free parameters are unknown, optimization is required to ensure the best fit for the data. In considering the effects of change trapping within our models, we established realistic bounds as priors for the charge trapping parameters. Assumed to be uniformly distributed, the initial slow trap population, having longer trapping lifetimes, was given a range of 0-500 counts, whereas the initial fast trap population, having shorter trapping lifetimes, was given a range of 0-200 counts. The combination of these priors on our parameters and the likelihood distribution from observed data returns a posterior distribution. Performing Markov Chain Monte Carlo (MCMC) analysis, as done in this study with the Python package \texttt{emcee} outlined in \citet{emcee}, allows one to estimate (sample) the posterior distribution to generatively model the data. Each random sample, generated by a sequential process, is used as a stepping stone to produce the next random sample (the chain). For each visit, we generated a chain for each of the ten wavelength channels. To ensure sufficiently converged chains (chains that have reached some equilibrium), we generated chains long enough to satisfy the recommended 50 times the integrated autocorrelation time for each parameter following methods in \citet{Goodman2010} as used in \texttt{emcee} \citep{emcee}. We ran each chain up to 20,000 steps. Using the autocorrelation times calculated from \texttt{emcee}, we established a burn-in and a thinning parameter. The burn-in—the early steps to be discarded because the chain has not yet converged—we defined as two times the parameter with the longest autocorrelation time. The thinning parameter thins the sample by every \textit{n}$^{th}$ step from the chain, and we defined it as half the parameter with the minimum autocorrelation time. Additionally, we flattened our 3D chains into 2D for plotting purposes. We used these flattened samples to create corner plots, 2D projections of the likelihood function \citep{foremanCorner}, as seen in Figure \ref{fig:cornerplot} of Appendix \ref{Corner Plots}. 

\section{Results and Discussion}
\label{Results_Discussion}

In this section, we present the emission and transmission spectra for exoplanet CoRoT-1 b and compare them against model emission spectra and previously published transmission spectra to constrain the atmospheric properties of this hot Jupiter. We generate a grid of self-consistent, 1D, radiative–convective–thermochemical equilibrium, cloud-free atmosphere models using the Self-consistent \textbf{C}altec\textbf{H} \textbf{I}nverse \textbf{M}od\textbf{E}ling and \textbf{R}etrieval \textbf{A}lgorithms (Sc-CHIMERA) code described in \citet{Piskorz2018} and \citet{arcangeli2018wasp18}. The original CHIMERA framework is outlined in \citet{line2013chemDiseq} and \citet{line2014CtOsecE}. Models utilize the two-stream approach \citep{toon1989} to calculate layer-by-layer fluxes from an internal source function and incident stellar flux using PHOENIX stellar spectral models \citep{husser2013}. A Newton–Raphson iteration method \citep{mckay1989} is used to achieve equilibrium vertical temperature profiles, and opacities are computed using the correlated-K method \citep{2017Amundsen}. Molecular abundances are determined with the NASA CEA2 \citep{gordon1994cea} Gibbs free energy minimization code, and condensate rainout is accounted for by depleting abundances in overlying layers \citep{1999BurrowsSharp}.

\subsection{Emission Spectra}
\label{Emission Spectra}

\begin{figure*}
\minipage{\textwidth}
    \includegraphics[width=\textwidth]{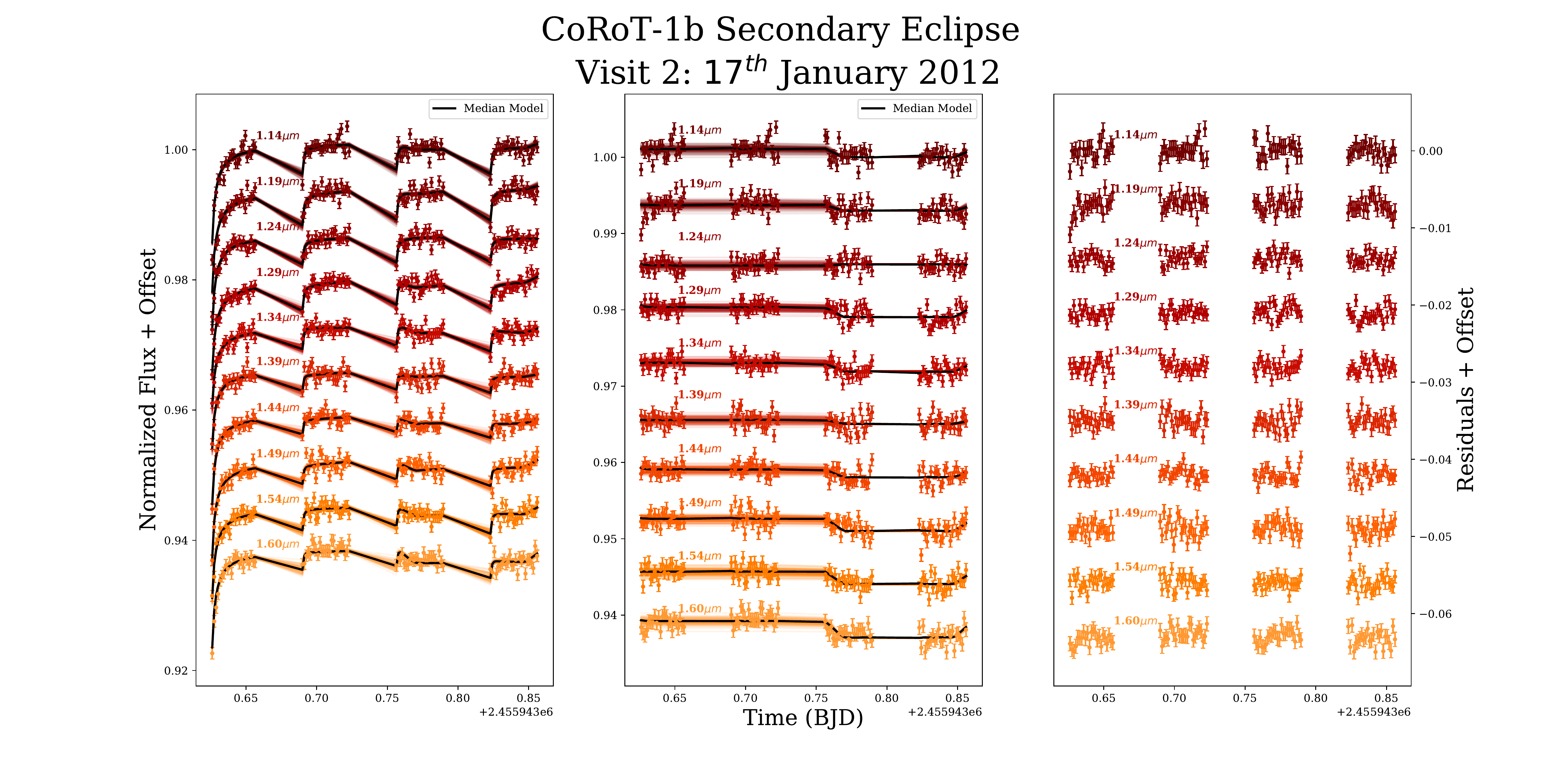}
   \caption{Secondary eclipse light curves of CoRoT-1 b, Visit 2. The first (leftmost) panel displays the raw light curves (the eclipse, baseline, and ramp profile) for the ten wavelength channels. One hundred randomly sampled light curves are drawn from the posterior distribution and overlaid with the median model shown as a solid black line. The second (middle) panel displays the astrophysical signal (the eclipse) with the ramp profile and baseline divided out of both the data and the randomly sampled light curves. The third (rightmost) panel displays the residuals from the median model.The residuals in any wavelength channel maximally varied by $\sim$ 1001 ppm from the median model. Additionally, Table \ref{tab:secondary_eclipse_parameters} notes the correlation coefficients between wavelengths. All panels display an applied flux offset in between wavelength channels for clarity. Notice here that the traditionally discarded first orbit remains included within these model fits.}
    \label{fig:lightcurve_visit2}
\endminipage
\vspace{0.5cm}
\minipage{\textwidth}
    \includegraphics[width=\textwidth]{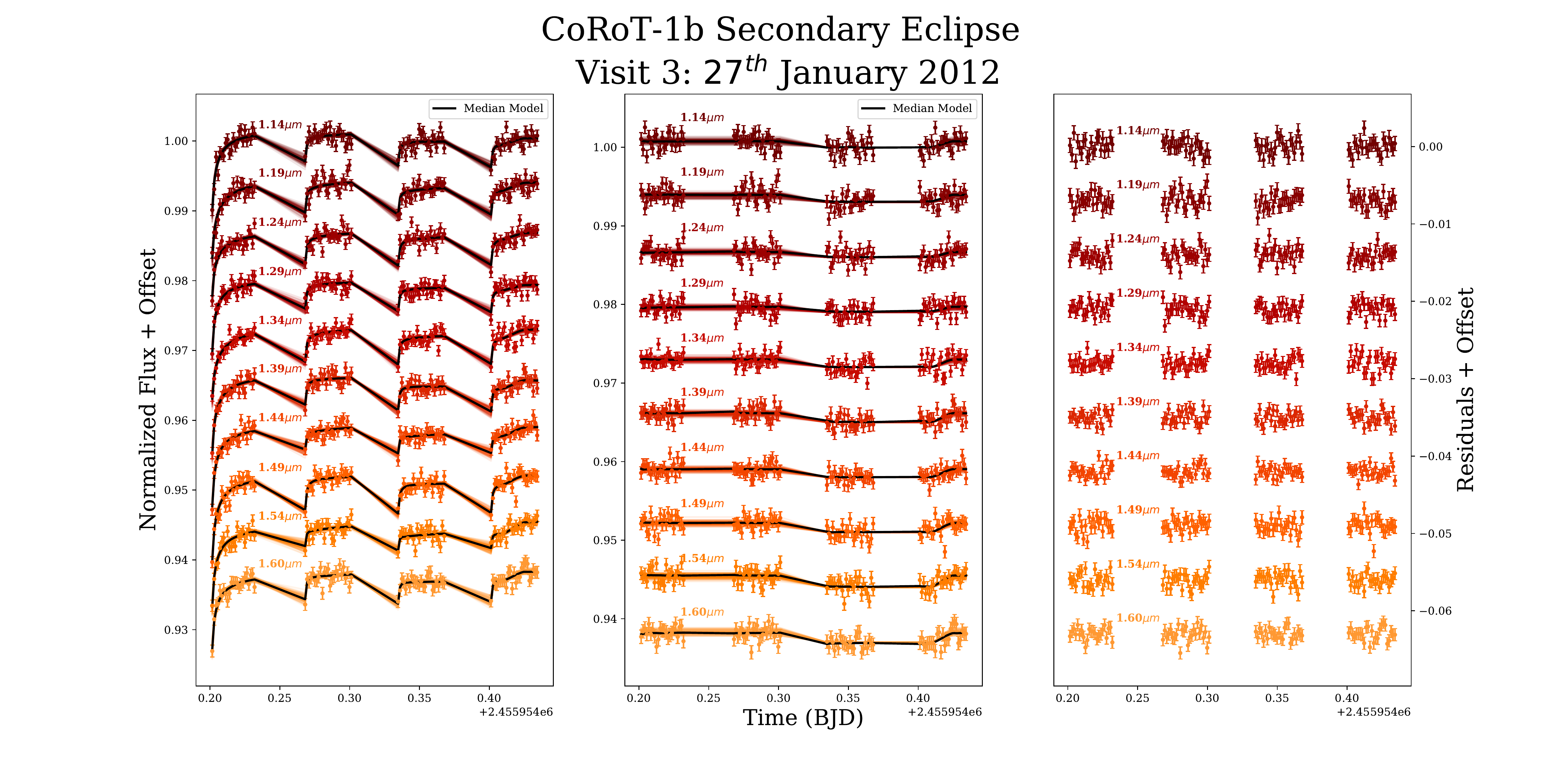}
    \caption{Secondary eclipse light curves of CoRoT-1 b, Visit 3. The residuals at any wavelength channel maximally varied by $\sim$ 924 ppm from the median model. Additionally, Table \ref{tab:secondary_eclipse_parameters} notes the correlation coefficients between wavelengths. Other comments are the same as for Figure \ref{fig:lightcurve_visit2}.
    \label{fig:lightcurve_visit3}}
\endminipage
\end{figure*}

\begin{figure*}
\minipage{\textwidth}
    \includegraphics[width=\textwidth]{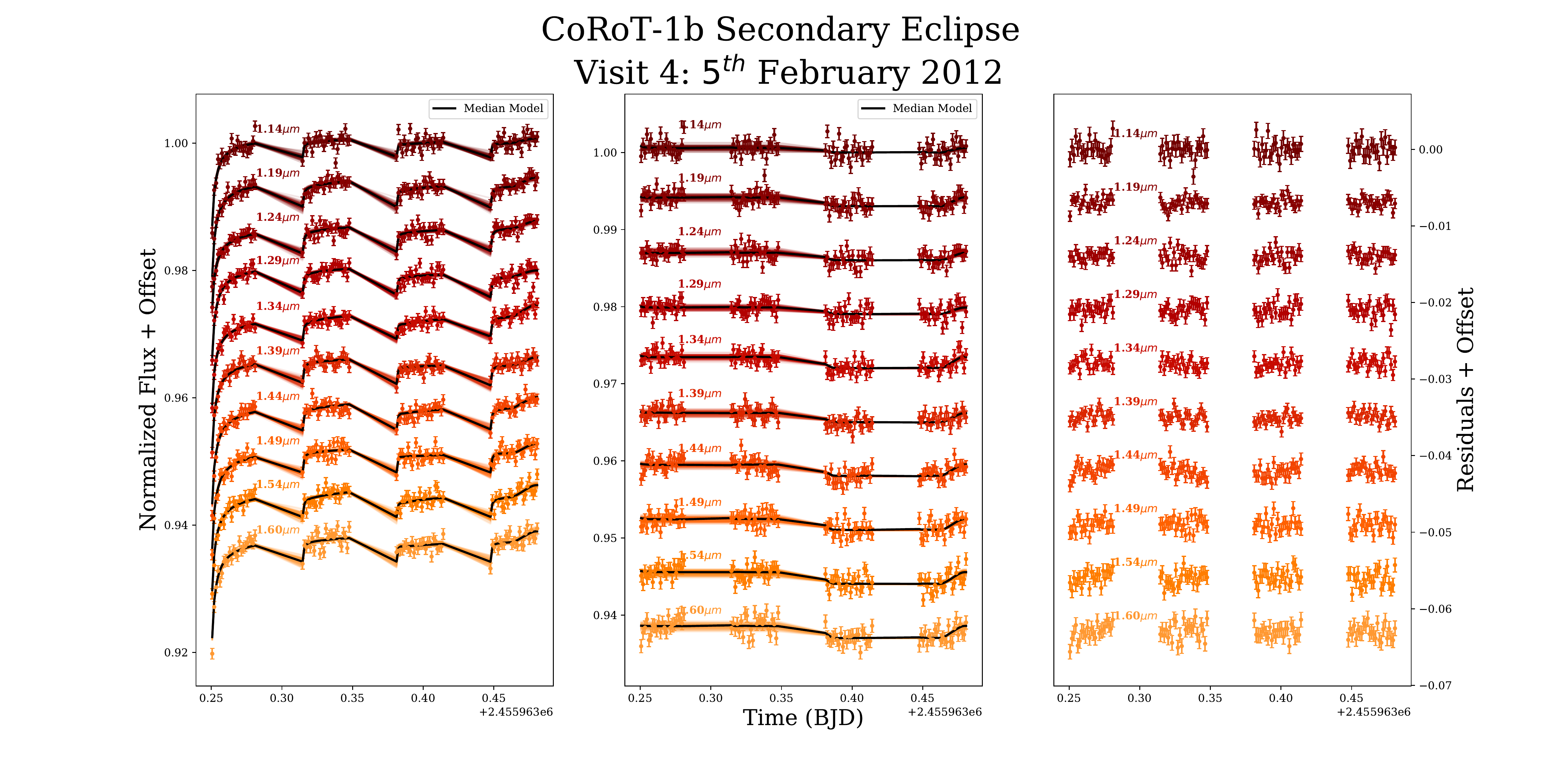}
   \caption{Secondary eclipse light curves of CoRoT-1 b, Visit 4. The residuals in any wavelength channel maximally varied by $\sim$ 989 ppm from the median model. Inspecting the residuals for Visit 4, the light curves appear to contain some correlated noise, particularly at the beginning of the first orbit. Referencing our Lomb–Scargle periodograms in Figure \ref{fig:lombscargle} of Appendix \ref{Lomb-Scargle Periodograms}, we conclude that the time correlation in the residuals is minor. Additionally, Table \ref{tab:secondary_eclipse_parameters} notes the correlation coefficients between wavelengths. Other comments are the same as for Figure \ref{fig:lightcurve_visit2}}. \label{fig:lightcurve_visit4}
\endminipage
\vspace{0.5cm}
\minipage{\textwidth}
    \begin{center}
    \includegraphics[width=0.7\textwidth,height=0.6\textwidth]{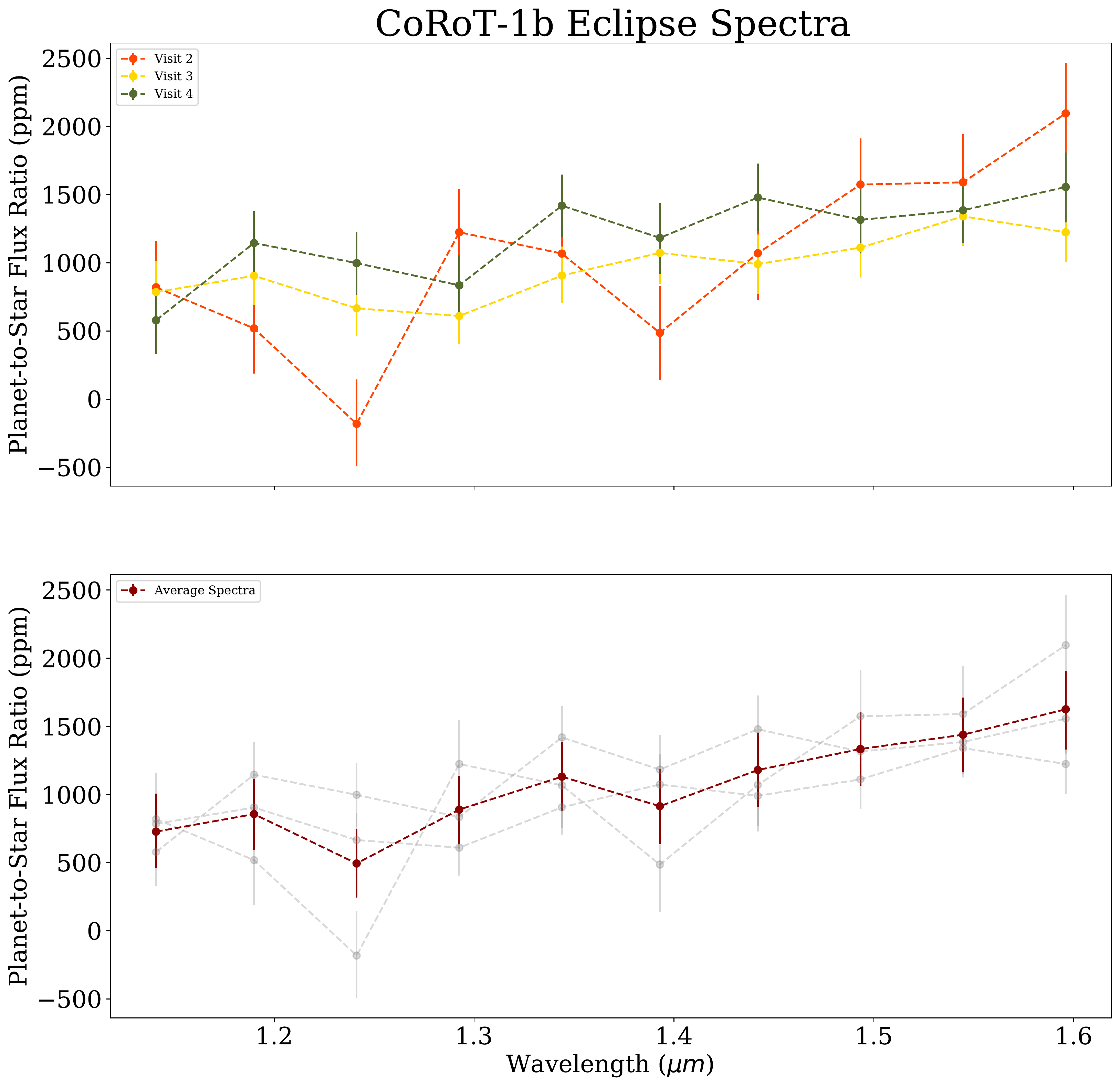}
    \caption{Top: individual emission spectra for each visit. Each emission spectrum is consistent with each other to within 1.4$\sigma$. Bottom: the averaged emission spectrum for CoRoT-1 b.
    \label{fig:emission_spectrum}}
    \end{center}
\endminipage
\end{figure*}

Figures \ref{fig:lightcurve_visit2}, \ref{fig:lightcurve_visit3}, and \ref{fig:lightcurve_visit4} present the secondary eclipse light curves with the optimized fits superimposed. For visualization purposes, the light curves, and their corresponding fits, are separated into three panels for raw data, corrected light curves, and residuals. Given observational uncertainties, MCMC returns random samples from the posterior distribution, each sample having an associated light curve. Considering 100 representative model fits, we determined the median model of these representative fits to calculate the residuals.

As seen in the upper panel of Figure \ref{fig:emission_spectrum}, each secondary eclipse visit was consistent with each other to within 1.4$\sigma$. Therefore, for our emission spectrum, we averaged the three secondary eclipse visits to increase signal-to-noise ratio. In addition to the averaged spectrum, we fit the three secondary eclipses simultaneously with a single model to check for consistency. Instead of modeling each visit separately, emergent light curves were phase-folded. In this method, each light curves requires correction for the ramp effect before phase-folding. We used the previously calculated data from our corrected light curves, as seen in the second panels of Figures \ref{fig:lightcurve_visit2}, \ref{fig:lightcurve_visit3}, and \ref{fig:lightcurve_visit4}, as they only contain the eclipse. We phase-folded the corrected light curves and performed an MCMC analysis to optimize the singular phase-folded eclipse model. To ensure the accuracy of our averaging method, we confirmed that the two methods agree well. Between the two methods, the planet-to-star flux ratio at any wavelength channel maximally varied by $\sim$ 211 ppm. Comparing posterior uncertainties, the largest error was $\sim$ 160 ppm, so the two methods are consistent within 1.3$\sigma$. Figure \ref{fig:emission_spectrum} presents the averaged emission spectrum from HST/WFC3 data on CoRoT-1 b. Table \ref{tab:secondary_eclipse_parameters} contains the best-fitting eclipse and charge trapping parameters in tabular form. 

\begin{deluxetable*}{ccccccc}
\tablenum{4}
\tablecaption{Optimized Secondary Eclipse Parameters}
\tablewidth{0pt}
\tablecolumns{12}
\tablehead{
\multicolumn{1}{c}{}& \multicolumn{1}{c}{Wavelength} & \multicolumn{1}{c}{Eclipse Depth$^*$} & \multicolumn{1}{c}{Slow Trap$^*$} &
\multicolumn{1}{c}{$\Delta$ Slow Trap$^*$} &
\multicolumn{1}{c}{Fast Trap$^*$} &
\multicolumn{1}{c}{$\Delta$ Fast Trap$^*$}\\
\multicolumn{1}{c}{Bin No.} & \multicolumn{1}{c}{($\micron$)} & \multicolumn{1}{c}{($F_{p}$/$F_{\star}$ $\times$ 10$^6$)} & 
\multicolumn{1}{c}{Population No.} &
\multicolumn{1}{c}{Population No.} &
\multicolumn{1}{c}{Population No.} &
\multicolumn{1}{c}{Population No.} 
}
\startdata
0 & 1.117–1.164 & $812_{-268}^{+269}$ & $111_{-68}^{+82}$ & $113_{-56}^{+70}$ & $11_{-6}^{+8}$ & $22_{-6}^{+7}$\\
1 & 1.164–1.215 & $921_{-257}^{+259}$ & $44_{-31}^{+56}$ & $51_{-5}^{+49}$ & $16_{-4}^{+6}$ & $19_{-12}^{+6}$\\
2 & 1.215–1.266 & $524_{-246}^{+256}$ & $216_{-88}^{+85}$ & $81_{-50}^{+67}$ & $21_{-8}^{+11}$ & $19_{-7}^{+7}$\\
3 & 1.266–1.318 & $919_{-247}^{+255}$ & $173_{-63}^{+66}$ & $65_{-43}^{+57}$ & $13_{-6}^{+10}$ & $22_{-7}^{+7}$\\
4 & 1.318–1.369 & $1185_{-247}^{+259}$ & $169_{-93}^{+116}$ & $143_{-58}^{+74}$ & $26_{-9}^{+10}$ & $19_{-6}^{+7}$\\
5 & 1.369–1.416 & $963_{-278}^{+280}$ & $149_{-96}^{+114}$ & $116_{-68}^{+79}$ & $19_{-9}^{+11}$ & $24_{-9}^{+9}$\\
6 & 1.416–1.467 & $1188_{-262}^{+265}$ & $125_{-71}^{+85}$ & $86_{-55}^{+71}$ & $11_{-6}^{+9}$ & $30_{-8}^{+8}$ \\
7 & 1.467–1.518 & $1388_{-271}^{+276}$ & $52_{-38}^{+65}$ & $148_{-74}^{+80}$ & $11_{-5}^{+7}$ & $17_{-8}^{+9}$\\
8 & 1.518–1.570 & $1487_{-280}^{+73}$ & $129_{-70}^{+81}$ & $112_{-56}^{+76}$ & $11_{-6}^{+9}$ & $27_{-7}^{+8}$\\
9 & 1.570–1.622 & $1672_{-296}^{+123}$ & $124_{-57}^{+53}$ & $186_{-93}^{+113}$ & $11_{-5}^{+7}$ & $21_{-9}^{+11}$
\enddata
\tablecomments{$^*$ The secondary eclipse optimized light curve fit parameters represent the average of the median models for all three secondary eclipse visits for each channel, respectively. Additionally, we calculated wavelength correlation coefficients between 0 and 0.3, extrapolating that a similar level of correlation may exist in the emission spectrum. \label{tab:secondary_eclipse_parameters}}\end{deluxetable*}

\begin{figure*}[htb!]
\begin{minipage}[c]{0.5\textwidth}
    \includegraphics[trim={1.5cm 2cm 0 3cm},clip,width=1.4\textwidth]{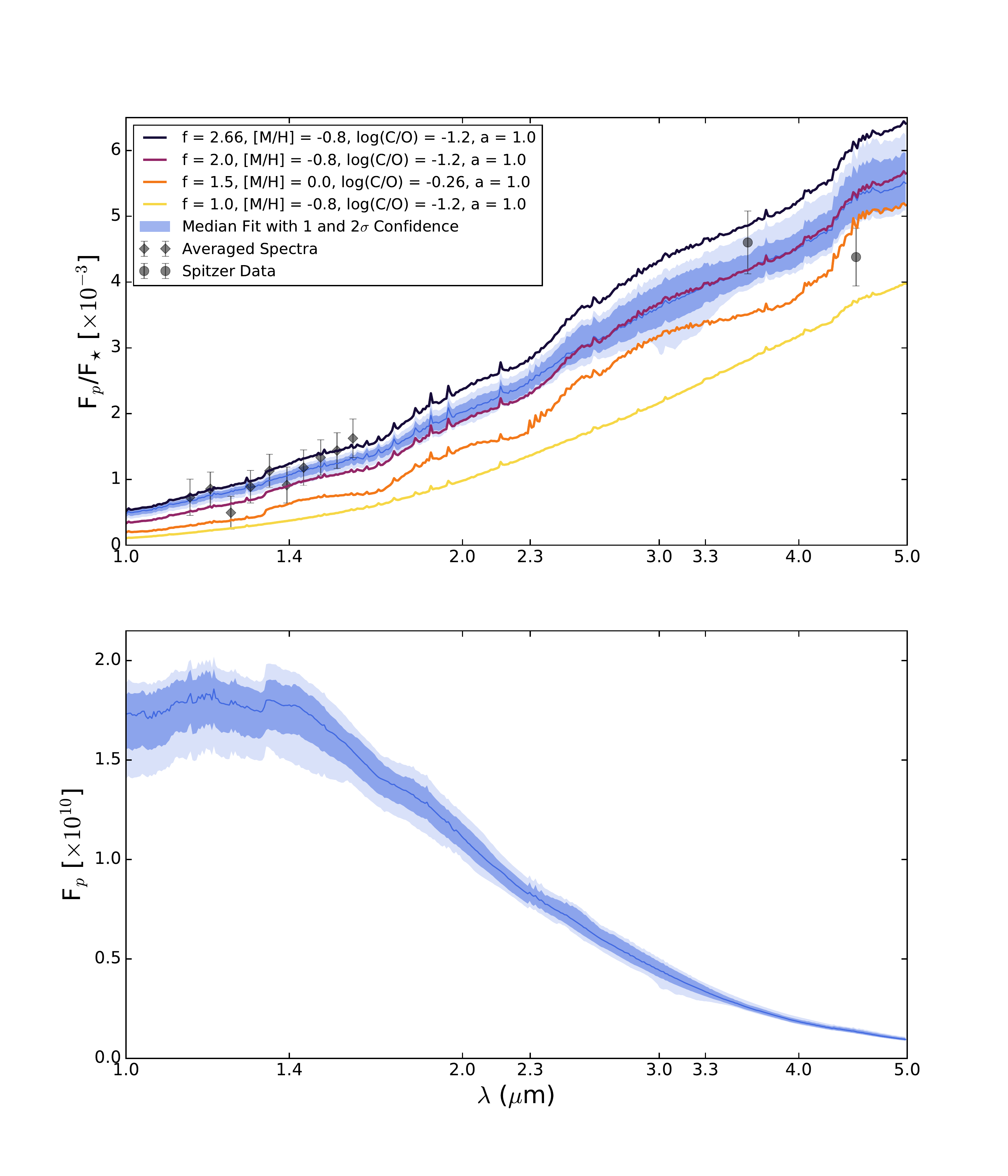}
\end{minipage}
\begin{minipage}[c]{0.3\textwidth}
\caption{Sc-CHIMERA modeled thermal emission spectra for CoRoT-1 b. The median fit closely corresponds to a retrieved maximum value of the heat redistribution parameter (\textit{f}), i.e., no redistribution, [M/H]=-0.8, log(C/O)=-1.2, and \textit{a}=0.87. However, constraints on [M/H] and log(C/O) are not significant. The scaling factor, \textit{a}, scales the resulting emission spectrum to account for a hot spot dominating the spectrum \citep{taylor2020}. Overlaid are four additional spectra with noted values for \textit{f}, [M/H], log(C/O), and \textit{a}. A heat redistribution parameter of 1.0 represents full redistribution, 2.0 represents dayside redistribution only, and 2.66 represents no heat redistribution with a concentrated hot spot at the substellar point. Finally, the points shown are the resulting average emission spectrum at HST wavelengths and corresponding Spitzer points at 3.6 and 4.5\micron. We updated Spitzer eclipse depths from \citet{deming11} as mentioned in section \ref{Spitzer Re-analysis}. In the bottom panel, the median fit has been multiplied by the stellar flux to isolate the planet flux and to highlight that the HST data probes the peak of the planet's flux. \label{fig:model_emission_spectrum}}
\end{minipage}
\begin{minipage}[b]{0.5\linewidth}
    \includegraphics[width=0.9\textwidth]{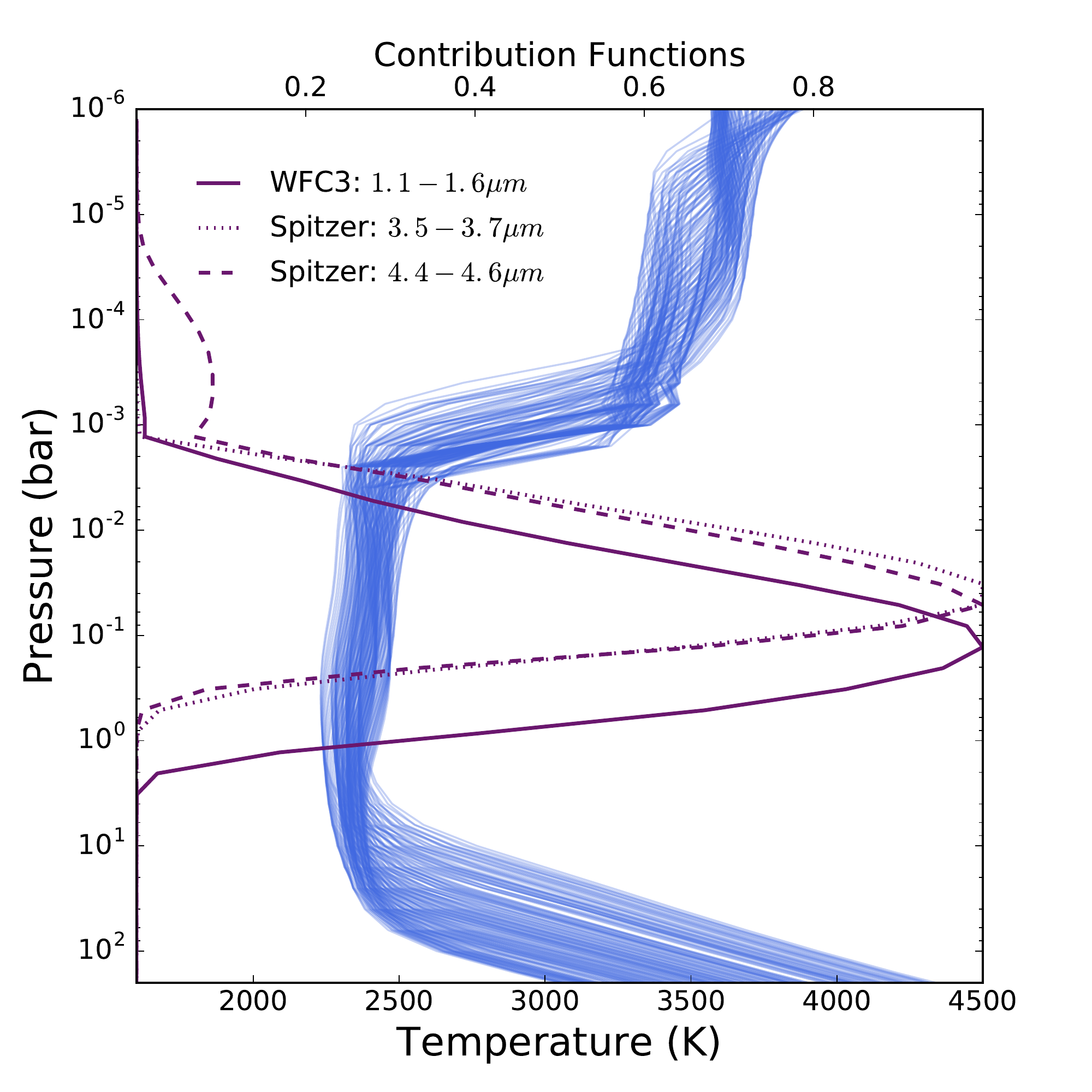}
    \caption{Sc-CHIMERA modeled pressure–temperature\\ profiles covering the $1 \sigma$ confidence range (Figure \ref{fig:grid_retrieval}). Also\\ plotted are contribution functions for the WFC3\\ wavelength range and each \textit{Spitzer} point. 
    \label{fig:TP_profile}}
\end{minipage} 
\begin{minipage}[b]{0.5\linewidth}
    \includegraphics[width=\textwidth]{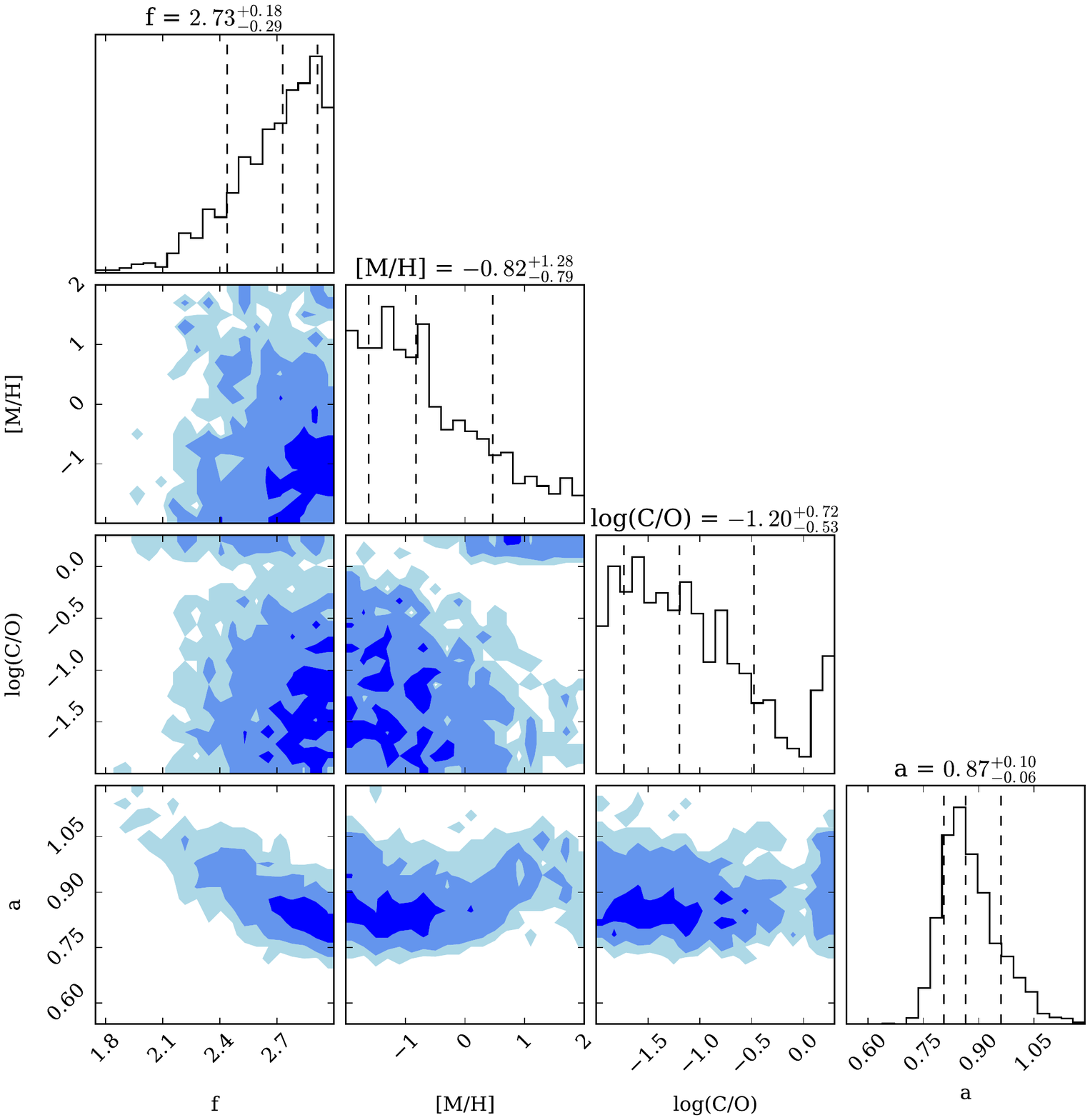}
    \caption{Posterior distributions from the grid retrieval. Models calculate log(C/O), [M/H], the heat redistribution parameter (\textit{f}), and the scaling factor (\textit{a}), which scales the resulting emission spectrum to account for a hot spot dominating the spectrum \citep{taylor2020}. Here, \textit{f} is consistent with no heat redistribution.
    \label{fig:grid_retrieval}}
\end{minipage}
\end{figure*}

Observing the emission spectrum we obtained for CoRoT-1 b, it appears to be generally featureless, similar to a blackbody-like emission as seen in Figure \ref{fig:emission_spectrum}. This HST/WFC3 emission spectrum does not exhibit any expected spectral features such as $H_2O$ or TiO. Sc-CHIMERA 1D grid modeling of the thermal emission spectrum for CoRoT-1 b suggests that the atmosphere’s best-fitting model has a heat redistribution parameter (defined in \citet{arcangeli2018wasp18}) \textit{f} = $2.73_{-0.29}^{+0.18}$ (see Figure \ref{fig:grid_retrieval}), indicating no heat redistribution, and contains a model-dependent thermal inversion (see Figure \ref{fig:model_emission_spectrum} \& \ref{fig:TP_profile}). Model constraints on the metallicity [M/H] = $-0.82_{-0.79}^{+1.28}$ and carbon-to-oxygen ratio log(C/O) = $1.2_ {-0.53}^{+0.72}$ (see Figure \ref{fig:grid_retrieval}) are not significant due to the absence of spectrally resolved molecular features. 

Thermal inversions are typically indicated by the presence of molecular emission bands. In similar HST/WFC3 studies, the ultra-hot Jupiter WASP-121 b (equilibrium temperature $\sim2500$ K) presented evidence for the presence of a stratosphere and $H_2O$ in emission, and the ultra-hot Jupiter WASP-33 b (equilibrium temperature $\sim~2700$ K) presented evidence for a temperature inversion caused by TiO \citep{2015WASP33b,WASP121bStrat}. Blackbody-like emission spectra have also shown up in similar HST/WFC3 studies. The ultra-hot Jupiters WASP-12 b and WASP-103 b (equilibrium temperatures $\sim2990$ K and $\sim2890$ K respectively) have returned blackbody-like emission spectra, indicating potential isothermal T–P profiles at the depths probed by HST wavelengths \citep{crossfield2012, 2017WASP103b}. WASP-103 b also suggests the idea of a pseudo-isothermal profile potentially explained by an overlaying thermal inversion layer or an upper atmosphere of clouds or haze \citep{2017WASP103b}. Additionally, a study with Spitzer on ultra-hot Jupiter WASP-18 b (equilibrium temperature $\sim2400K$) has also presented a potential isothermal T–P profile. Also, Spitzer's analysis of WASP-12 b suggests the molecular absorption of CO in the $4.5\micron$ channel, despite the reported blackbody-like spectrum in the HST bandpass \citep{crossfield2012}.  

Now, delving deeper into emissions, within the near-IR atmospheric emission between hot Jupiters and ultra-hot Jupiters, a transition between inverted and non-inverted thermal profiles, predicted in \citet{mansfield2018hatp7}, has been reported around an equilibrium temperature of 1700 K (assuming zero albedo and full redistribution) by \citet{baxter2020}. The Spitzer IRAC-based transition at 1700 K was reported based on deviations from a blackbody, defined by the brightness temperature measured from eclipse depths at $3.6$ \micron, from observed $4.5$ $\micron$ eclipse depths, with greater observed deviations indicating stronger thermal inversions \citep{baxter2020}. The  1700 K equilibrium temperature transition, interestingly enough, corresponds to the condensation temperatures of TiO and VO, supporting the idea that the presence of these high-altitude absorbers can potentially introduce temperature inversions to the T–P profile  \citep{hubeny03,fortney08twoclass,baxter2020}. Even in small amounts, these species are strong absorbers capable of heating the upper atmosphere of these hot Jupiters \citep{kreidberg2018review}. These inversions can also be a result of other absorbers in the upper atmosphere. According to \citet{schlawin2014}, based on their Infrared Telescope Facility (IRTF) spectrum and CoRoT-1 b photometry for the primary transit, strong absorption caused by TiO/VO is disfavored but, high-altitude hazes could affect the transmission spectrum. In this paper, however, our Sc-CHIMERA modeled thermal emission spectrum is cloud-free. \citet{baxter2020} reported CoRoT-1 b to have an equilibrium temperature equal to 1900 K (assuming zero albedo and full redistribution), supporting what we calculated in Table \ref{tab:fixedparams}, which places CoRoT-1 b, initially, in a transitional realm between hot Jupiters and ultra-hot Jupiters. This transitional realm poses the possibility of observing a temperature inversion alongside corresponding spectral features. However, evidence of low heat redistribution efficiencies on CoRoT-1 b has been indicated and could suggest an atmosphere exhibiting different behaviors from its cooler counterparts (\textit{T}$<$1500 K) \citep{fortney08twoclass,Zhao_2011,baxter2020}.

Previously published results from the IR wavelength range have already found CoRoT-1 b to be consistent with a blackbody, a spectrum not typically indicative of a thermal inversion, based on measurements at 3.6 and 4.5 $\micron$ with the Spitzer IRAC \citep{deming11,baxter2020}. Specifically, \citet{baxter2020} references Spitzer eclipse depth values of  \textit{F}$_p$/\textit{F}$_\star$ = 4150$\pm$420 ppm and \textit{F}$_p$/\textit{F}$_\star$ = 4820$\pm$420 ppm for 3.6 and 4.5 $\micron$ respectively on CoRoT-1 b from \citet{deming11}. \citet{deming11} reported that CoRoT-1 b was best reproduced by a 2460 K blackbody, while \citet{baxter2020} reported brightness temperatures between 2200 and 2300K. These reported temperatures support our Sc-CHIMERA modeled thermal emission spectrum corresponding to no heat redistribution, with a calculated equilibrium temperature of 2450K (assuming zero albedo and no heat redistribution; see Table \ref{tab:fixedparams}). Lower efficiency in heat redistribution would return higher equilibrium temperatures on the dayside as well as higher brightness temperatures. The low efficiency is an expectation from circulation models, which predict that radiative cooling causes large day and night temperature differences for planets with \textit{T} $>$ 2000 K \citep{komacek2017circulation}. Calculated equilibrium temperatures for various redistribution efficiencies are in Table \ref{tab:fixedparams}. 
Although blackbody-like spectra do not typically indicate thermal inversions, thermal inversions are highly likely in high-temperature bodies, and the blackbody-like spectra produced are likely due to molecular dissociation of water as well as H$^-$ opacity. Dominant opacity sources in the near-IR for hot Jupiters include $H_2O$ and CO. In ultra-hot Jupiters with temperatures around 2500 K, water begins to thermally dissociate on the dayside and becomes depleted beyond 2700 K \citep{arcangeli2018wasp18, parmentier2018phaseCurveReview}. As water dissociates, opacities from hydrogen ions become dominant on the daysides of these hot Jupiters \citep{arcangeli2018wasp18, parmentier2018phaseCurveReview}. On the other hand, CO, a strongly bonded molecule, is difficult to dissociate, and according to \citet{arcangeli2018wasp18} it should be present below 4000 K. Therefore, a temperature inversion should result in observable CO emission \citep{baxter2020}. We begin to see this emission in the upper wavelength range toward the Spitzer 4.5 $\micron$ channel of our model of the thermal emission spectrum (Figure \ref{fig:model_emission_spectrum}). We note that a more recent pipeline and analysis (see section \ref{Spitzer Re-analysis}) returns Spitzer eclipse depth values of  \textit{F}$_p$/\textit{F}$_\star$ = 4602$\pm$477 ppm and \textit{F}$_p$/\textit{F}$_\star$ = 4380$\pm$440 ppm for 3.6 and 4.5 $\micron$ respectively for CoRoT-1 b. Figure \ref{fig:model_emission_spectrum} reflects these updated values. Noticeably different from the previously reported wavelength values is the flip in the 3.6 and 4.5 $\micron$ values from \citet{deming11}. Contrary to the emission in the model, the new reduction could indicate CO absorption at 4.5 $\micron$; however, given the emission spectrum and the T–P of the planet, a CO absorption could be challenging. Given that this is from a single point in the spectrum we do not draw any firm conclusions about the CO feature. A spectrum from JWST, which has the capacity to span the wavelength range from 0.6 $\micron$ to 28 $\micron$, could provide more insight.
\subsection{Transmission Spectra}
\label{sec:Transmission Spectra}

\begin{figure*}[htb!]
\minipage{\textwidth}
    \includegraphics[width=\textwidth]{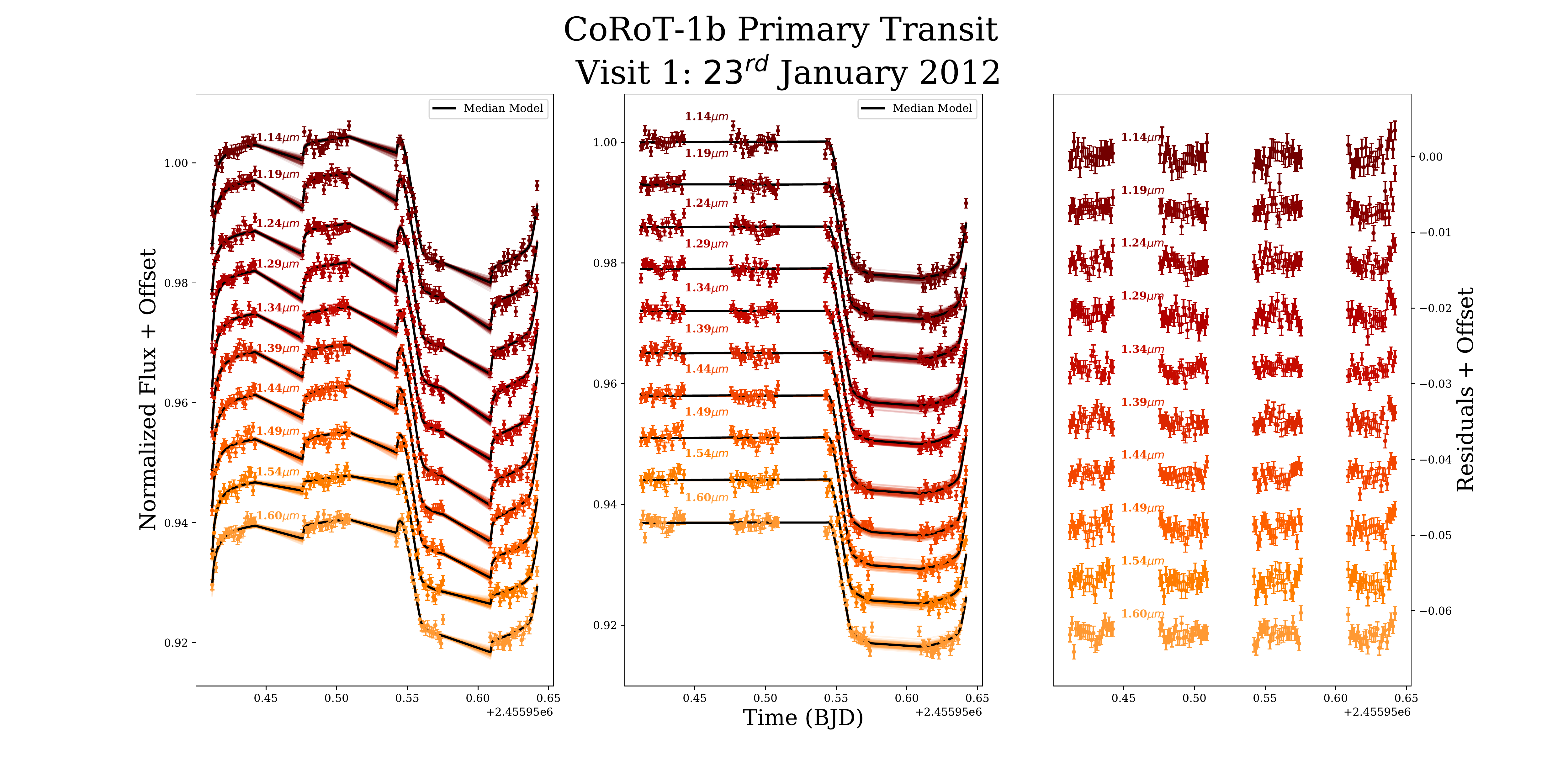}
   \caption{Primary transit light curves of CoRoT-1 b, Visit 1. The residuals in any wavelength channel maximally varied by $\sim$ 1247 ppm from the median model. Additionally, Table \ref{tab:primary_eclipse_parameters} notes the correlation coefficients between wavelengths. Other comments are the same as for Figure \ref{fig:lightcurve_visit2}.} 
   \label{fig:lightcurve_visit1}
\endminipage
\vspace{0.5cm}
\minipage{\textwidth}
    \begin{center}
    \includegraphics[width=\textwidth]{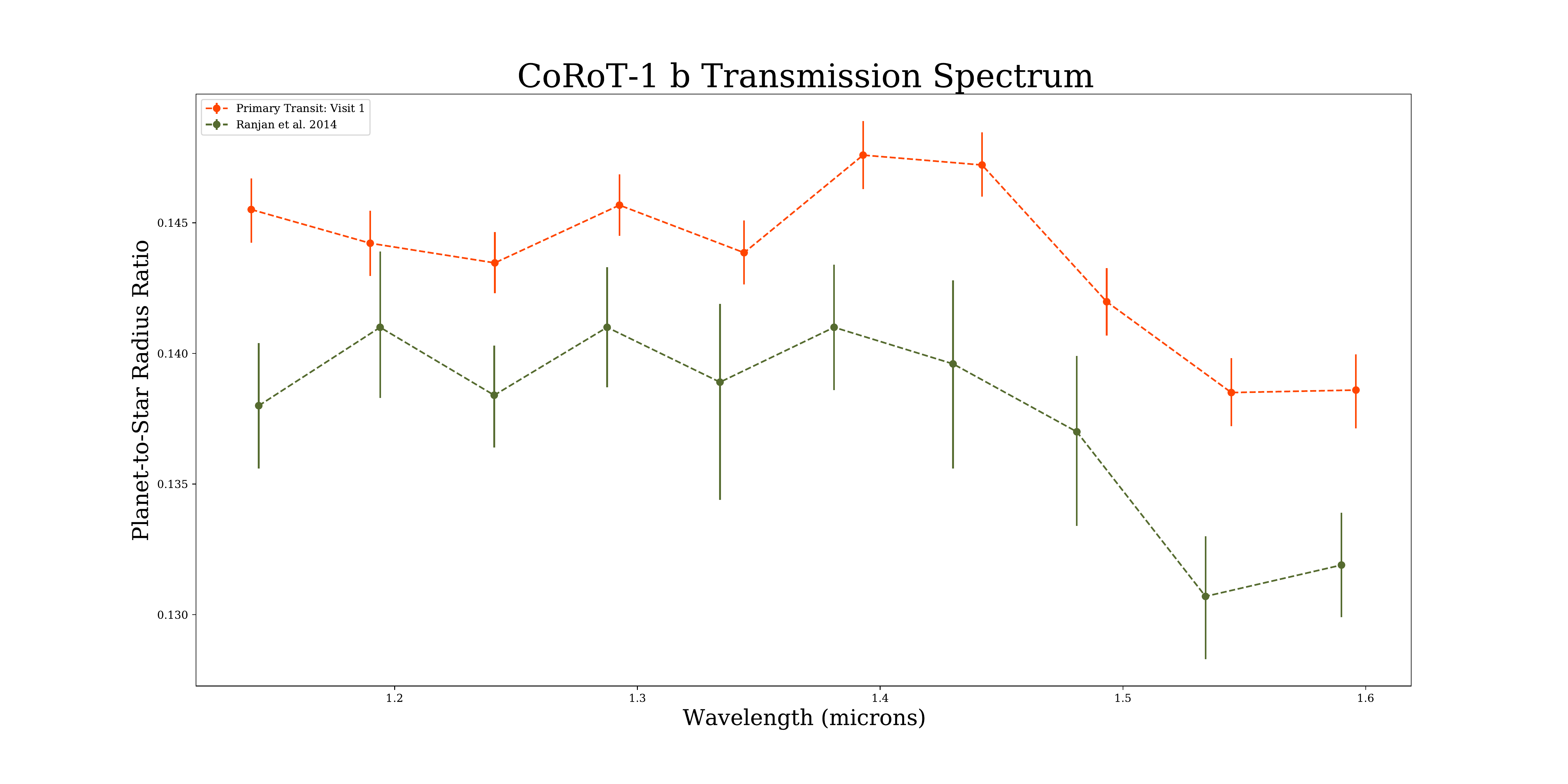}
    \caption{Comparison of CoRoT-1 b transmission spectrum with results from \citet{ranjan2014ApJ...785..148R}. 
    \label{fig:transmission_spectrum}}
    \end{center}
\endminipage
\end{figure*}

In addition to the HST/WFC3 emission spectra, we briefly analyzed the single primary transit, previously studied and published. Figure \ref{fig:lightcurve_visit1} presents the corresponding light curve with optimized fits superimposed. The same analysis process as with the secondary eclipse light curves, was done for the primary transit light curve, with the exception of a limb darkening model. We assumed the same four-parameter limb darkening profile as in \citet{ranjan2014ApJ...785..148R} with coefficient values [0.396, 0.571, -0.748, 0.286]. We pulled the required values from \citet{ranjan2014ApJ...785..148R} and used the Exoplanet Characterization Toolkit (ExoCTK) Limb Darkening Calculator to obtain the profile. Figure \ref{fig:transmission_spectrum} presents the transmission spectrum from HST/WFC3 data on CoRoT-1 b. Table \ref{tab:primary_eclipse_parameters} contains best-fit spectra and the charge trapping parameters in tabular form. In the first report of primary transit detection of exoplanet CoRoT-1 b, \citep{barge08} reported \textit{R}$_p$/\textit{R}$_\star$ = 0.1388$\pm$0.0021. However, throughout the literature this value has fluctuated, with near-UV \textit{R}$_p$/\textit{R}$_\star$ = $0.1439_{+0.0020}^{-0.0018}$ \citep{turner2016MNRAS.459..789T}, optical \textit{R}$_p$/\textit{R}$_\star$ = $0.1381_{+0.0007}^{-0.0015}$ \citep{gillon09}, infrared \textit{R}$_p$/\textit{R}$_\star$ = 0.147 $\pm$ 0.002 (weighted average) \citep{schlawin2014}, and \textit{R}$_p$/\textit{R}$_\star$ = 0.1433 $\pm$ 0.0010 \citep{bean09}. In this study, we compared the planet-to-star radius ratios for each wavelength bin (found in Table \ref{tab:primary_eclipse_parameters}) to results from \citet{ranjan2014ApJ...785..148R}. Figure \ref{fig:transmission_spectrum} presents the resulting and the comparison transmission spectra. It is clear there is an offset between the two sets of data, and even if different aperture sizes are used, CoRoT-1 b does not appear to have any companions nearby causing transit depth discrepancies according to the HST direct image. There are also no companions reported in \citet{evans2016A&A...589A..58E}. Therefore, we can only assume the offset is due to discrepancies in the detector systematic corrections or pipelines, which also caused a 0.0045 \textit{R}$_p$/\textit{R}$_\star$ offset between \citet{bean09} and \citet{barge08} (both observed at the same wavelength). One specific difference between the two analyses done is the use of the divide-out-of-transit method \citep{berta2012flat_gj1214} to correct for the prominent ramp effect systematic in \citet{ranjan2014ApJ...785..148R}. We implemented \texttt{RECTE} to account for this systematic in our results. Despite the offset and different method of analysis, our transmission spectrum also supports conclusions reflected in our comparison with \citet{ranjan2014ApJ...785..148R})—a generally featureless spectrum, consistent with a flat line. We also see the same small drop in \textit{R}$_p$/\textit{R}$_\star$ toward 1.55 $\micron$, which may have to do with the partial transit, i.e. no baseline following the transit.

\begin{deluxetable*}{ccccccc}
\tablenum{5}
\tablecaption{Optimized Primary Eclipse Parameters \label{tab:primary_eclipse_parameters}}
\tablewidth{0pt}
\tablecolumns{12}
\tablehead{
\multicolumn{1}{c}{}& \multicolumn{1}{c}{Wavelength} & \multicolumn{1}{c}{Planet-to-Star Radius Ratio} &
\multicolumn{1}{c}{Slow Trap} &
\multicolumn{1}{c}{$\Delta$ Slow Trap} &
\multicolumn{1}{c}{Fast Trap} &
\multicolumn{1}{c}{$\Delta$ Fast Trap}\\
\multicolumn{1}{c}{Bin No.} & \multicolumn{1}{c}{($\micron$)} & \multicolumn{1}{c}{($R_{p}$/$R_{\star}$)} & 
\multicolumn{1}{c}{Population No.} &
\multicolumn{1}{c}{Population No.} &
\multicolumn{1}{c}{Population No.} &
\multicolumn{1}{c}{Population No.} 
}
\startdata
 0 & 1.117–1.164 & $0.1455_{-0.0012}^{+0.0012}$ & $443_{-81}^{+42}$ & $13_{-9}^{+22}$ & $17_{-8}^{+11}$ & $31_{-5}^{+6}$\\
 1 & 1.164–1.215 & $0.1442_{-0.0012}^{+0.0013}$ & $367_{-129}^{+88}$ & $19_{-13}^{+30}$ & $17_{-10}^{+14}$ & $10_{-5}^{+6}$\\
 2 & 1.215–1.266 & $0.1435_{-0.0011}^{+0.0012}$ & $415_{-107}^{+62}$ & $36_{-25}^{+48}$ & $24_{-10}^{+12}$ & $14_{-6}^{+6}$\\
 3 & 1.266–1.318 & $0.1457_{-0.0011}^{+0.0012}$ & $449_{-65}^{+38}$ & $17_{-12}^{+25}$ & $6_{-4}^{+7}$ & $6_{-3}^{+5}$\\
 4 & 1.318–1.369 & $0.1439_{-0.0012}^{+0.0013}$ & $341_{-95}^{+84}$ & $21_{-15}^{+32}$ & $5_{-3}^{+8}$ & $15_{-5}^{+6}$\\
 5 & 1.369–1.416 & $0.1476_{-0.0013}^{+0.0014}$ & $218_{-103}^{+103}$ & $39_{-28}^{+49}$ & $7_{-4}^{+9}$ & $11_{-5}^{+6}$\\
6 & 1.416–1.467 & $0.1472_{-0.0012}^{+0.0013}$ & $252_{-110}^{+101}$ & $18_{-13}^{+29}$ & $7_{-5}^{+10}$ & $16_{-5}^{+6}$ \\
7 & 1.467–1.518 & $0.1420_{-0.0012}^{+0.0013}$ & 
$392_{-109}^{+73}$ & $17_{-12}^{+27}$ & $11_{-7}^{+11}$ & $24_{-5}^{+6}$\\
8 & 1.518–1.570 & $0.1385_{-0.0012}^{+0.0014}$ & $256_{-123}^{+117}$ & $14_{-10}^{+22}$ & $12_{-8}^{+13}$ & $48_{-5}^{+6}$\\
9 & 1.570–1.622 & $0.1360_{-0.0014}^{+0.0014}$ & $245_{-145}^{+141}$ & $23_{-16}^{+34}$ & $40_{-15}^{+16}$ & $39_{-6}^{+7}$
\enddata
\tablecomments{All values pulled from the median model for each wavelength bin. Additionally, we calculated wavelength correlation coefficients between 0 and 0.3, extrapolating that a similar level of correlation may exist in the transmission spectrum.
}\end{deluxetable*}

\section{Conclusion}
\label{sec:Conclusion}
We derived an emission spectrum and a transmission spectrum for exoplanet CoRoT-1 b using the WFC3 instrument aboard the HST. Contrary to popular dividing-out-of-transit methods, which discard the first orbit's flux data due to the ramp effect, we generated light curve models implementing \texttt{RECTE}'s physically motivated solutions.  As a new generation of observations with JWST approaches, which will also host stare mode, a better understanding of the effects of systematics is critical for characterizing exoplanets.

The HST/WFC3 emission and transmission spectra both indicate that CoRoT-1 b is a featureless blackbody. According to Sc-CHIMERA models of thermal emission spectra, the blackbody-like spectrum corresponds best to an atmosphere with no heat redistribution. We calculated the equilibrium temperature (assuming zero albedo and no heat redistribution) as 2450K. We note from \citet{garhart2020} that the common practice is to estimate temperatures of hot Jupiter assuming zero albedo and a uniform heat redistribution. However, based on a statistical characterization of hot Jupiter atmospheres using Spitzer secondary eclipses, \citet{garhart2020} found that uniform dayside redistribution was a more accurate reflection of the temperature. Applying CoRoT-1 b's degree of heat redistribution to this larger population of planets \citep{garhart2020}, we note that it falls outside of the median results of the study. 
We consider the possibility that the blackbody-like spectrum results from an isothermal atmosphere, which is supported in the pressure region probed by the WFC3 but, based on Sc-CHIMERA models, CoRoT-1 b returns an inverted T–P profile. Inverted T–P profiles are predictively indicative of an absorber in the upper layer of the atmosphere. However, based on the featureless spectra and the temperature of CoRoT-1 b, we assume that the thermal dissociation of water and H$^-$ opacity is responsible. As our spectra did not return any features, further investigation will be needed to constrain the atmospheric properties of CoRoT-1 b. While we expect to see the dissociation of water, CO is an emission feature we expect in this temperature regime. Looking forward, we expect the JWST to probe new pressure levels in and around the CO bands where deviations from a blackbody will be more apparent, providing stronger evidence for these thermal inversions. 
These observations and analysis pave the way to high-precision results with stare mode spectroscopy. The time-series spectroscopic modes of JWST's NIRSpec, MIRI, and NIRCam instruments will be analogous to HST's stare mode used in this study. 

\section*{Acknowledgements}
Thanks to Rafia Bushra, whose thesis provided a valuable springboard for the results of the paper.
Funding for E. Schlawin and K. Glidic is provided by NASA Goddard Space Flight Center. 
We respectfully acknowledge the University of Arizona is on the land and territories of Indigenous peoples. Today, Arizona is home to 22 federally recognized tribes, with Tucson being home to the O'odham and the Yaqui. Committed to diversity and inclusion, the University strives to build sustainable relationships with sovereign Native Nations and Indigenous communities through education offerings, partnerships, and community service.

\vspace{0.5cm}
\facilities{HST(WFC3)}

\software{\texttt{astropy} \citep{astropy2013},
\texttt{batman} \citep{batman},
\texttt{corner} \citep{foremanCorner},
\texttt{emcee} \citep{emcee},
\texttt{matplotlib} \citep{Hunter2007matplotlib},
\texttt{numpy} \citep{vanderWalt2011numpy},
\texttt{RECTE} \citep{zhou2017chargeTrap},
\texttt{scipy} \citep{virtanen2020scipy},
\texttt{tshirt} (\url{https://tshirt.readthedocs.io/en/latest/})}

\appendix
As we refer to a few plots and their implications to our results without directly inserting them, we take this opportunity to provide additional commentary on the aforementioned figures.
\counterwithin{figure}{section}
\vspace{0.5cm}

\section{Corner Plots}
\label{Corner Plots} 

The corner plot, first mentioned in section \ref{sec:Optimizing Light Curve Fits}, is used to visualize the results of our MCMC light curve fits. In a corner plot, the posterior probability distribution of each parameter is projected in a one- and two-dimensional space. 1D histograms are along the diagonal, representing the marginalized distribution for each parameter. The remaining panels display the marginalized 2D distributions \citep{ emcee,foremanCorner}. The 2D distributions demonstrate the covariance between parameters (how the different parameters vary together). Figure \ref{fig:cornerplot} of Appendix \ref{Corner Plots} shows a representative corner plot with the seven parameters considered in our models.

\begin{figure*}[htb!]
\begin{center}
\includegraphics[scale=0.3]{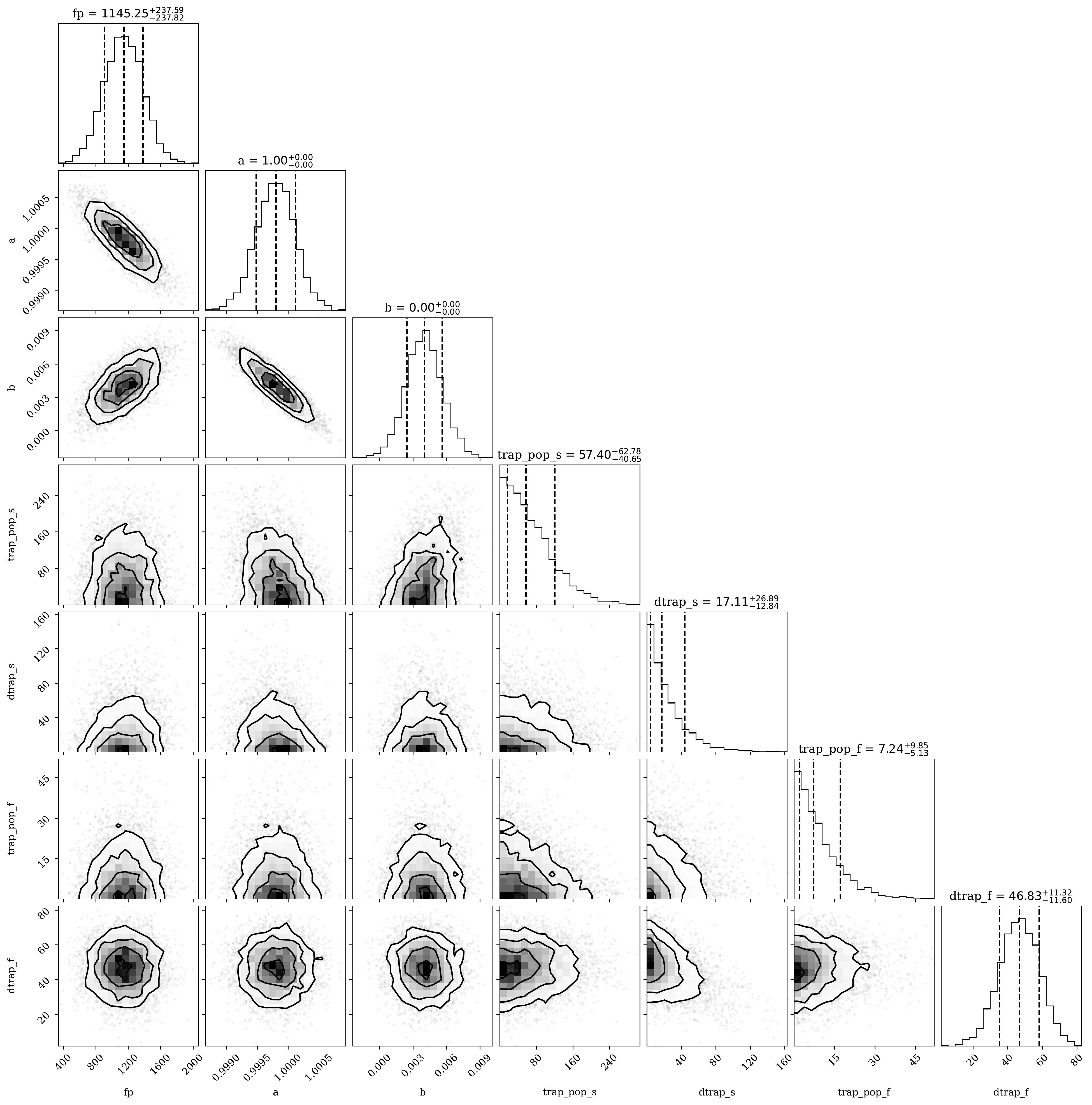}
\caption{Representative secondary eclipse model corner plot of the posterior distributions pulled from Visit 4, wavelength channel 2 (1.164–1.215 $\micron$). The parameter names are defined as used in the Python script. Parameter \texttt{fp} is the planet-to-star flux ratio. Parameter \texttt{a} and \texttt{b} are the \textit{y}-intercept and the slope of the linear regression we applied to the models respectively. The four \texttt{RECTE} charge trap parameters are: the initial number of occupied charge carrier traps for the fast and slow populations \texttt{(trap\_pop\_s, trap\_pop\_f)} and the number of extra trapped charge carriers added in the middle of two orbits \texttt{(dtrap\_s, dtrap\_f)} for each population respectively.In the 1D histograms, the middle dashed line represents the mean result while the two adjacent lines represent $1\sigma$ from the mean.
\label{fig:cornerplot}}
\end{center}
\end{figure*}

\clearpage

\section{Lomb-Scargle Periodograms}
\label{Lomb-Scargle Periodograms}
The Lomb–Scargle periodogram in Figure \ref{fig:lombscargle} of Appendix \ref{Lomb-Scargle Periodograms}, first mentioned in Figure \ref{fig:lightcurve_visit4}, is used to support our conclusion that the time correlation in our residuals is minor. Lomb–Scargle periodograms compute Fourier-like power spectra that detect and characterize periodic signals in unevenly sampled data \citep{lomb, scargle, lombscargle}. In Figure \ref{fig:lombscargle} we utilized the \texttt{astropy LombScargle} function to specifically search for periodic signatures and any other deviations from white noise in our light curve residuals.

\begin{figure*}[htb!]
\begin{center}
\includegraphics[scale=0.3]{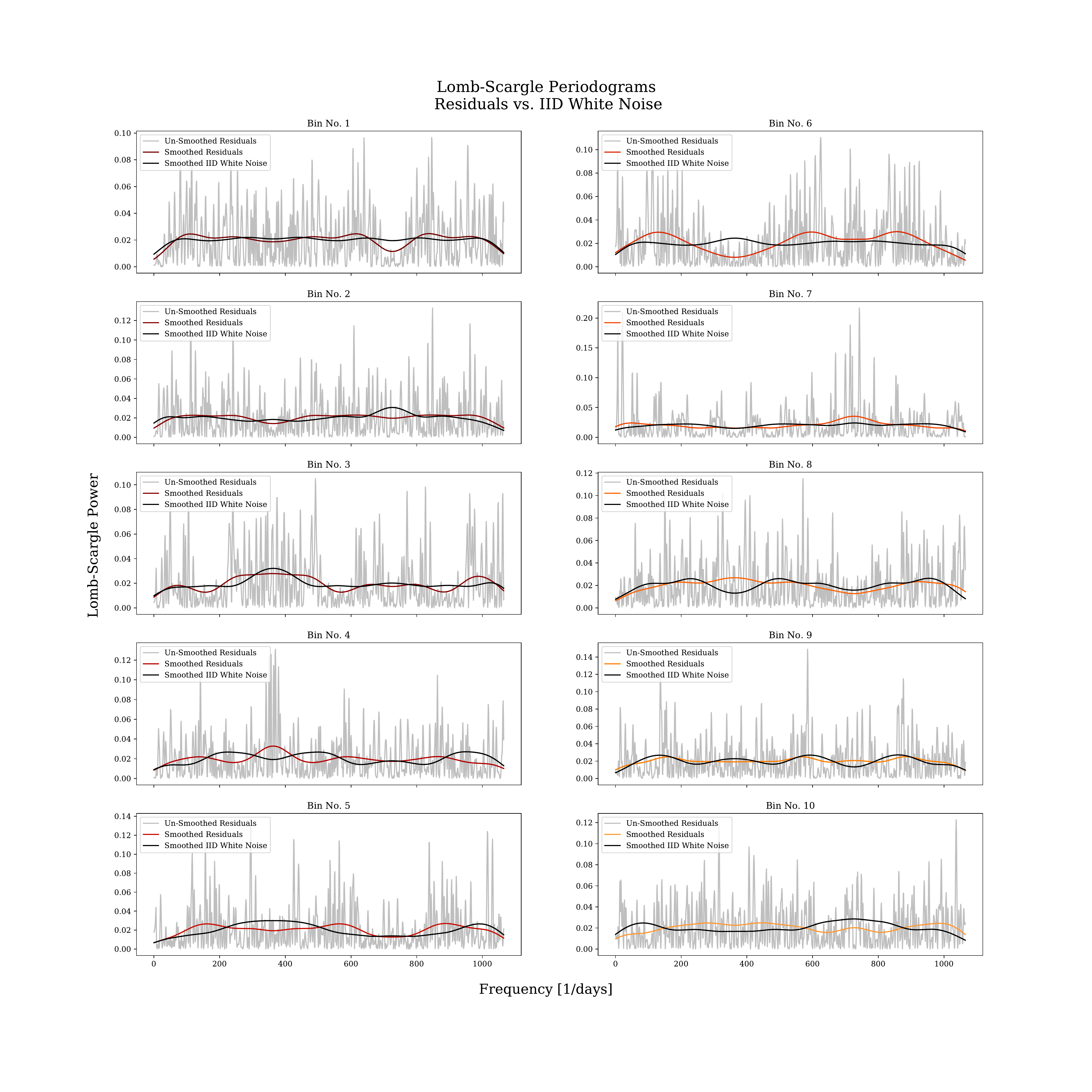}
\caption{Lomb–Scargle residual periodograms for each wavelength channel in Visit 4. Additionally overlaid are the smoothed residual and independently and identically distributed (IID) white noise periodograms, done with the \texttt{astropy} 1D Gaussian filter kernel with a standard deviation of 50. The smoothed residual and IID white noise periodograms show a slight statistical deviation that, overall, supports very minor residual time-correlated noise in the light curves of Figure \ref{fig:lightcurve_visit4}. 
\label{fig:lombscargle}}
\end{center}
\end{figure*}

\clearpage
\bibliography{CoRoT-1_b}{}
\bibliographystyle{aasjournal}

\end{document}